\documentclass[nohyper,12pt,letterpaper]{JHEP3}

\usepackage{amsmath,amssymb,amsfonts,xspace}
\usepackage[latin1]{inputenc}
\usepackage{multirow}

\numberwithin{equation}{section}

\newcommand{\de}{\mathrm{d}}
\newcommand{\De}{\mathrm{D}}
\newcommand{\I}{\mathrm{i}}
\newcommand{\e}{\mathrm{e}}
\newcommand{\cL}{\mathcal{L}}

\newcommand{\p}{\partial}

\newcommand{\half}{\frac{1}{2}}
\newcommand{\qtr}{\frac{1}{4}}

\newcommand{\cC}{\mathcal{C}}
\newcommand{\cS}{\mathcal{S}}
\newcommand{\cG}{\mathcal{G}}
\newcommand{\cK}{\mathcal{K}}
\newcommand{\cM}{\mathcal{M}}
\newcommand{\cN}{\mathcal{N}}
\newcommand{\cE}{\mathcal{E}}
\newcommand{\cX}{\mathcal{X}}

\newcommand{\cT}{\mathcal{T}}
\newcommand{\cJ}{\mathcal{J}}

\newcommand{\La}{\Lambda}
\newcommand{\Si}{\Sigma}

\newcommand{\zb}{\bar{z}}

\DeclareSymbolFont{AMSa}{U}{msa}{m}{n}
\DeclareSymbolFont{AMSb}{U}{msb}{m}{n}
\DeclareMathSymbol{\fieldR}{\mathalpha}{AMSb}{"52}

\newcommand{\kahler}{{K\"ahler}\xspace}
\newcommand{\hk}{{hyperk\"ahler}\xspace}
\newcommand{\qk}{{quaternionic-K\"ahler}\xspace}
\newcommand{\C}{{\mathbb C}}
\newcommand{\PP}{{\mathbb P}}
\newcommand{\R}{{\mathbb R}}
\newcommand{\Z}{{\mathbb Z}}
\newcommand{\cZ}{\mathcal{Z}}
\newcommand{\cO}{\mathcal{O}}
\newcommand{\abs}[1]{\lvert#1\rvert}
\newcommand{\norm}[1]{\lVert#1\rVert}
\newcommand{\cech}{{\v{C}ech}\xspace}

\newcommand{\tC}{{\tilde C}}
\newcommand{\ltr}[1]{\left\langle #1 \right\rangle}
\newcommand{\inprod}[1]{\langle#1\rangle}
\newcommand{\sympquotient}{/\!/}

\renewcommand{\Im}{{\rm Im}}
\renewcommand{\Re}{{\rm Re}}

\newcommand{\pa}{\partial}
\newcommand{\nn}{\nonumber}
\newcommand{\cc}{\mathrm{c.c.}}
\newcommand{\eps}{\epsilon}
\newcommand{\CN}{{\cal N}}
\newcommand{\IR}{\mathbb{R}}
\newcommand{\IC}{\mathbb{C}}

\newcommand{\IP}{\mathbb{P}}

\newcommand{\tzeta}{\tilde\zeta}
\newcommand{\txi}{\tilde\xi}

\def\bea{\begin{eqnarray}}
\def\eea{\end{eqnarray}}
\def\be{\begin{equation}}
\def\ee{\end{equation}}
\def\ba{\begin{align}}
\def\ea{\end{align}}
\def\bse{\begin{subequations}}
\def\ese{\end{subequations}}

\title{Twistors and Black Holes}

\preprint{\hepth{0701214}\\LPTENS-07-02\\
ITP-UU-07-02\\SPIN-07-02}

\author{Andrew Neitzke\footnote{Email: {\tt  neitzke@ias.edu, pioline@lpthe.jussieu.fr, vandoren@phys.uu.nl}}\\
School of Natural Sciences, Institute for Advanced Study, Princeton, NJ, USA}

\author{Boris Pioline$^*$\\
-- 
Laboratoire de Physique Th\'eorique et Hautes Energies\footnote{Unit\'e mixte
de recherche du CNRS UMR 7589}, \\
Universit\'e Pierre et Marie Curie - Paris 6,
4 place Jussieu, F-75252 Paris cedex 05 \\
-- Laboratoire de Physique Th\'eorique de l'Ecole Normale 
Sup\'erieure\footnote{Unit\'e mixte
de recherche du CNRS UMR 8549}\\
24 rue Lhomond, F-75231 Paris cedex 05}

\author{Stefan Vandoren$^*$ \\ Institute for Theoretical Physics 
{\rm and} Spinoza Institute\\
Utrecht University, 3508 TD Utrecht, The Netherlands}

\abstract{Motivated by black hole physics
in $\cN=2, D=4$ supergravity, we study the geometry of quaternionic-K\"ahler 
manifolds $\cM$ obtained by the $c$-map construction from projective special
K\"ahler manifolds $\cM_s$.
Improving on earlier treatments, 
we compute the K\"ahler potentials on the twistor space $\cZ$ and
Swann space $\cS$ in the complex coordinates adapted to the Heisenberg 
symmetries. The results bear a simple relation to the Hesse potential $\Sigma$ of the
special K\"ahler manifold $\cM_s$, and hence 
to the Bekenstein-Hawking entropy for BPS black holes.
We explicitly construct the ``covariant $c$-map'' and the ``twistor map'',
which relate real
coordinates on $\cM \times \IC\IP^1$ (resp. $\cM \times \R^4 / \Z_2$)
to complex coordinates on $\cZ$ (resp. $\cS$).
As applications, we solve for the general BPS geodesic motion on $\cM$,
and provide explicit integral formulae for the quaternionic Penrose
transform relating elements of $H^1(\cZ, {\cal O}(-k))$ to massless fields
on $\cM$ annihilated by first or second order differential operators.
Finally, we compute the exact radial wave function (in the supergravity
approximation) for BPS black holes with fixed
electric and magnetic charges.}

\begin{document}


\section{Introduction}
In this paper, we study the geometry of \qk manifolds $\cM$ obtained by the 
$c$-map construction of \cite{Cecotti:1988qn,Ferrara:1989ik} from a
projective special K\"ahler manifold $\cM_s$. While our results may
be of independent mathematical interest, our motivation stems from
the physics of BPS black holes in $\cN=2, D=4$ supergravity, as we now
explain. The mathematically oriented or impatient reader is kindly urged to proceed
to Section \ref{sec-outline}.

\subsection{Motivation}

Supersymmetric black holes in Type II string theory compactified on
a Calabi-Yau three-fold $X$ offer a rich playground to test the
stringy description of black-hole micro-states beyond
leading order: on the macroscopic side, thanks to an off-shell
superspace description of vector multiplets, an infinite series of
higher-derivative curvature corrections can be computed using the
topological string on $X$ \cite{Bershadsky:1994cx, Antoniadis:1994ze};
on the microscopic side,
the weakly coupled D-brane \cite{Strominger:1996sh, Maldacena:1996gb}
or M5-brane \cite{Maldacena:1997de} description can be extended to
strong coupling, thanks to the tree-level decoupling
between vector multiplets and hypermultiplets.
The interplay between these two descriptions has culminated in a recent conjecture
\cite{Ooguri:2004zv} relating the microscopic degeneracies, to all
orders in an expansion in the inverse of the graviphoton charge,
to the topological string amplitude (see \textit{e.g.}
\cite{Pioline:2006ni} for a recent review and further references).

Due to the aforementioned decoupling between vector multiplets and
hypermultiplets,
the study of BPS black holes in $\CN=2$ supergravity is usually framed
in the language of special geometry.  It has however become increasingly
clear that quaternionic-K\"ahler geometry may be a more useful
framework.  Indeed, the attractor equations that govern the radial evolution
of the complex vector multiplet scalars in the black hole geometry are
equivalent to ``supersymmetric'' geodesic motion on a para-quaternionic-K\"ahler manifold $\cM^*$,
of dimension $4n$ (where $n-1$ is the number of vector multiplets)
and split signature \cite{Gunaydin:2005mx}.  
This $\cM^*$ is a particular analytic continuation
(studied in \cite{Gunaydin:2005mx,Cortes:2005uq}) of the (positive signature)
quaternionic-K\"ahler manifold $\cM$, obtained via the $c$-map construction of
\cite{Cecotti:1988qn,Ferrara:1989ik} from the projective
special K\"ahler manifold
$\cM_s$ describing the vector multiplet scalars
in four dimensions.  This description of the attractor equations follows from
the fact that stationary black holes in four dimensions can be reduced
to three dimensions along their timelike isometry, where they
become solutions of three-dimensional Euclidean gravity coupled
to a non-linear sigma model on $\cM^*$.  Further
assuming spherical symmetry leads to geodesic motion on $\cM^*$
\cite{Breitenlohner:1987dg}; the electric, magnetic and NUT charges
of the black hole are identified as conserved Noether charges
for a Heisenberg algebra of isometries of $\cM^*$.
This equivalence between black hole attractor equations and geodesic
motion on a quaternionic-K\"ahler manifold can also be seen as a
consequence of T-duality along the time direction, which relates black holes
to D-instantons, with a non-trivial radial dependence of the hypermultiplets
in the dual four-dimensional theory \cite{Behrndt:1997ch,Gutperle:2000ve,
deVroome:2006xu}.

This reformulation is particularly well suited to the radial
quantization of BPS black holes, which, according to the
proposal in \cite{Ooguri:2005vr}, should provide a holographic
point of view on the conjecture of \cite{Ooguri:2004zv}.
Indeed, once the radial evolution equations are reformulated as
geodesic motion, quantization could proceed as usual
by replacing functions on the classical phase space $T^*(\cM^*)$,
of real dimension $8n$, by (square integrable) wave functions on $\cM$,
satisfying the appropriate Wheeler-De Witt type constraint
\cite{Gunaydin:2005mx}.  The corresponding Hilbert space is infinite
dimensional, even after restricting to the subspace with fixed electric and
magnetic charges.

More relevant however is the quantization
of supersymmetric geodesic motion, corresponding to BPS
black holes: the analysis in \cite{gnpw-to-appear}
(as announced in \cite{Gunaydin:2005mx,Pioline:2006ni}) 
shows that after imposing the BPS constraints the classical phase space 
becomes the 
twistor space $\cZ$ of $\cM$, with real dimension $4n+2$,
almost twice as small as the non-BPS phase space. The twistor
space is a standard construct in quaternionic
geometry \cite{MR664330}, which carries a K\"ahler-Einstein metric and a
canonical integrable complex structure, unlike the base $\cM$
whose quaternionic structure has a non-vanishing Nijenhuis
tensor.
It is fibered by 2-spheres over $\cM$; physically,
the coordinate in the fiber keeps track of the projectivized Killing
spinor preserved by the black hole \cite{gnpw-to-appear}.
In fact, it is convenient to integrate the entire quaternionic
structure on $\cM$ by introducing an $\R^4 / \Z_2$ bundle $\cS$
over the quaternionic-K\"ahler space $\cM$, known as the
the ``hyperk\"ahler cone'' or ``Swann space'' \cite{MR1096180};
the twistor space $\cZ$ then arises as a K\"ahler quotient 
${\cS} \sympquotient U(1)$. 
This construction is particularly natural in the conformal approach to
supergravity \cite{deWit:2001dj}, and leads to a simple
description of the supersymmetric geodesics in terms of holomorphic
maps from $\IC$ to $\cS$ \cite{gnpw-to-appear}.

Having identified the twistor space $\cZ$ as the BPS phase space,
quantization proceeds in the usual way for K\"ahler
manifolds, \textit{i.e.} by replacing functions on $\cZ$ with
classes in the cohomology of an appropriate line bundle over $\cZ$.
In more mundane terms, the BPS Hilbert space consists of
holomorphic functions in $2n+1$ variables.  In stark
contrast to the non-BPS case, the BPS wave function
is now uniquely specified by the electric and magnetic charges
of the black hole, as a coherent eigenstate of the Heisenberg 
symmetries \cite{gnpw-to-appear}.
It can be pushed down to a wave function on the base space $\cM$, 
annihilated by the quantum BPS constraints, by contour 
integration along the $\IC\IP^1$
fiber (a quaternionic generalization of the Penrose 
transform, described in \cite{quatman,MR1165872}.)

While the above statements hold on very general grounds,
for practical purposes it is important to have a direct handle on the
geometry of the twistor space $\cZ$ and the Swann space ${\cS}$.
In particular, it is necessary to know the K\"ahler potential on $\cZ$
explicitly, since it controls the inner product on the BPS Hilbert space.
To compute the Penrose transform of the BPS wave function
on $\cZ$, one also needs to express the complex coordinates
on $\cZ$ and $\cS$ in terms of the real coordinates on $\cM$
arising from the $c$-map and the complex coordinates on the fibers.
This lays the groundwork for a forthcoming study
of the radial quantization of BPS black holes \cite{gnpw-to-appear},
and possibly other physical applications.

\subsection{Outline\label{sec-outline}}
With this motivation in mind, the goal of the present work
is to further elucidate the geometry of the twistor space $\cZ$ and the
Swann space ${\cS}$, and in particular obtain explicit formulae for their
respective complex structures and K\"ahler potentials.

The outline of this paper is as follows. In Section 2, we review 
the construction of the twistor space $\cZ$ and Swann space $\cS$ on a
general quaternionic-K\"ahler manifold $\cM$; their description
in terms of a ``generalized prepotential'' $G$ (not to be confused with the
one controlling higher-derivative corrections on the vector multiplet side)
in cases when sufficiently many tri-holomorphic isometries are present;
and the relation, recently found in \cite{Rocek:2005ij,Rocek:2006xb},  
between $G$ and the prepotential $F$ 
in the case when $\cM$ is the $c$-map of a projective
special K\"ahler manifold $\cM_s$.

In Section 3, we compute the hyperk\"ahler potential $\chi$ on $\cS$
and the K\"ahler potential $K_{\cZ}$ on $\cZ$, by relaxing the
$SU(2)$ gauge choice made in  \cite{Rocek:2005ij}; the latter was sufficient
for the purpose of computing the metric on the quaternionic-K\"ahler base
but unsuitable for our present purposes.
In particular, we uncover a simple relation \eqref{kz-hesse}
between the K\"ahler potential
$K_\cZ$ and the Hesse potential $\Sigma$ on $\cM_s$, or equivalently the
Bekenstein-Hawking entropy of four-dimensional BPS black
holes. We also construct the ``covariant $c$-map'' \eqref{covcmap}
and ``twistor map'' \eqref{gentwi},
which relate the complex coordinates on $\cS$ or $\cZ$, adapted
to the Heisenberg symmetries, to the real
coordinates on $\cM \times (\IR^4 / \Z_2)$
or $\cM \times \IC\IP^1$, respectively.

In Section 4, we apply these techniques to find the general solution
for BPS geodesic motion on the $c$-map manifold $\cM$; this is relevant
to the problem of constructing spherically symmetric BPS black holes
or instantons in $\cN=2$ supergravity.  While the physical results obtained are
not new, this exercise illustrates the power of the twistor formalism, uncovers
the algebraic geometry behind these BPS configurations, and provides a physical
explanation for the relation between $K_\cZ$ and the black hole entropy.

In Section 5, we use the twistor map found in Section 3 to give a fully
explicit integral representation \eqref{twistor-ci-Ok-cmap} 
of the quaternionic Penrose transform,
which relates elements of $H^1(\cZ,\cO(-k))$ to functions on
the quaternionic-K\"ahler base $\cM$, satisfying certain massless field
equations. We also find the relevant inner product \eqref{inner}, under some
assumptions that we spell out.
As an example, we compute the
Penrose transform \eqref{bpswf} of an eigenmode \eqref{cohst}
of the Heisenberg group with vanishing central character on $\cZ$. 
As will be argued in \cite{gnpw-to-appear}, this is the
exact radial wave function for a BPS black hole with fixed
electric and magnetic charges, in the two-derivative supergravity
approximation.

Finally, some additional formulae and derivations used in the main text
are given in an Appendix at the end of this paper.

Many of the results in this paper were first observed by studying
$c$-maps based on Hermitian symmetric tube domains.
As we preview in Section
\ref{quartic}, the corresponding twistor spaces provide a transparent
geometric realization of certain group representations
constructed in \cite{Gunaydin:2000xr},
which will be discussed in a separate paper \cite{gnpw-to-appear-2}.

\subsection{Notation\label{sec-notation}}

For the reader's convenience, we collect here and in  Table \ref{tablenot}
some notation used (and defined) at various places throughout the paper.
Throughout the paper $\cM$ is a \qk manifold of real dimension $4n$.  
Except in Sections \ref{twiswa}, \ref{3holg}  and \ref{penrose-general},  
$\cM$ is obtained by the $c$-map from 
a special \kahler manifold $\cM_s$. The table summarizes the various spaces 
related to $\cM$ and the coordinate systems used in the paper. The range 
of the indices are $a \in \{1, \dots, n-1\}$, $\Lambda \in \{0\} \cup \{a\} 
= \{0, \dots, n-1\}$, $I \in \{\flat\} \cup \{\Lambda\} =
\{\flat, 0, \dots, n-1\}$, $A' \in \{1, 2\}$, $A \in \{1, \dots, 2n\}$.  
We use generic coordinates $x^\mu$ on $\cM$ to lighten the notation in 
statements not depending on a 
particular coordinate system; similarly $(u^i, \bar{u}^i)$ and $(z^\aleph, 
\bar{z}^{\bar \aleph})$ 
denote generic complex coordinates for $\cZ$ and $\cS$.
We sometimes drop indices inside arguments of functions, \textit{e.g.} 
we write the \kahler potential on $\cZ$ as $K_\cZ (u, \bar{u})$.
We emphasize that $z, z^a, z^\aleph$ are all unrelated, as are $x^I, x^\mu$ and
$\zeta, \zeta^\Lambda$ ($\zeta$ is a coordinate on the twistor space of $\cS$, 
introduced in Section \ref{3holg}).

\TABLE{
\begin{tabular}{|c|c|c|c|} \hline
Notation  &  Space  &  Real dim  &  Coordinate systems  \\ \hline \hline
$\cM_s$  &  special \kahler manifold  &  $2n-2$   &   $(z^a, \bar{z}^a)$ \\ 
\hline 
\multirow{2}{*}{$\cM$}  &  \qk manifold & \multirow{2}{*}{$4n$}  &  $(U, \zeta^\Lambda, \tzeta_\Lambda, \sigma, z^a, \bar{z}^a)$ \\
& ($c$-map of $\cM_s$)  & & $(x^\mu)$ \\
\hline
\multirow{3}{*}{$\cZ$} & \multirow{3}{*}{
\begin{tabular}{c}
complex contact manifold \\ 
(twistor space of $\cM$)
\end{tabular}
} & \multirow{3}{*}{$4n+2$} & $(\xi^\Lambda, \txi_\Lambda, \alpha, \bar{\xi}^\Lambda, \bar{\txi}_\Lambda, \bar{\alpha})$ \\ 
& & & $(x^\mu, z, \bar{z})$ \\
& & & $(u^i, \bar{u}^i)$ \\
\hline
\multirow{2}{*}{$\cJ$} & 3-Sasakian manifold & \multirow{2}{*}{$4n+3$} & \multirow{2}{*}{$(u^i, \bar{u}^i, \phi)$} \\
& ($S^3$ bundle over $\cM$) & & \\
\hline
\multirow{5}{*}{$\cS$} & & \multirow{5}{*}{$4n+4$} & $(v^I, \bar{v}^I, x^I, \theta_I)$ \\
& \multirow{3}{*}{
\begin{tabular}{c}
\hk manifold \\ 
(Swann space of $\cM$)
\end{tabular}
} & & $(v^I, \bar{v}^I, w_I, \bar{w}_I)$ \\
& & & $(u^i, \lambda, \bar{u}^i, \bar{\lambda})$ \\
& & & $(x^\mu, \pi^{A'}, \bar{\pi}^{A'})$ \\
& & & $(z^\aleph, \bar{z}^{\bar{\aleph}})$ \\
\hline
\end{tabular}
\caption{Overview of the manifolds discussed in the paper and their coordinate
systems.\label{tablenot}}
}

\section{Projective superspace description of the $c$-map: a review \label{review}}

In this section, we review the projective superfield description
of the $c$-map, first obtained in \cite{Rocek:2005ij,Rocek:2006xb}.
In Section \ref{twiswa}, we recall some standard facts about
the geometry of quaternionic K\"ahler manifolds, their Swann space and 
twistor space. 
In Section \ref{3holg}, we review the construction of the metric
on ${\cS}$ in the tensor multiplet formalism, in the case where
${\cS}$ admits $n+1$ commuting triholomorphic isometries.
Finally, in Section \ref{cmapg}, we specialize to the case where $\cM$ is obtained via the
$c$-map from a special K\"ahler manifold.  In this case we review the relation between the ``generalized
prepotential'' $G$ entering the tensor multiplet construction and the prepotential $F$
on the special K\"ahler base.

\subsection{Geometry of quaternionic-\kahler manifolds}\label{twiswa}

In this subsection, we collect some standard results about
the geometry of quaternionic-K\"ahler manifolds. Most of these
facts can be found in \cite{MR664330,MR1096180} 
or inferred from these references.

A quaternionic-K\"ahler manifold $\cM$ is a Riemannian manifold of real
dimension $4n$ with holonomy group $USp(2n) USp(2) = (USp(2n) \times USp(2)) / \Z_2$.
The complexified tangent bundle of such an $\cM$ splits locally as
\be
T_{\IC} \cM = E \otimes H\,,
\ee
where $E$ and $H$ are complex vector bundles of respective dimensions $2n$ and $2$.
This decomposition is preserved by the Levi-Civita connection.
Hence after choosing local frames for $E$ and $H$
one can trade the vector index $\mu$ in $T_{\IC} \cM$ for 
a pair of indices $AA'$, where
$A \in \{1, \dots, 2n\}$ and $A' \in \{1, 2\}$.  Concretely this is accomplished by
contracting with the ``quaternionic vielbein'',
a covariantly constant matrix of one-forms $V^{AA'}=V^{AA'}_\mu {\rm d}x^\mu$.
We will sometimes convert between $\mu$ and $AA'$ without writing $V$ explicitly.
$V$ satisfies a pseudo-reality condition
\be
(V^{AA'})^* = \eps_{AB} \eps_{A'B'} V^{BB'}\,.
\ee
Here $\eps_{AB} = \eps_{[AB]}$ and $\eps_{A'B'} = \eps_{[A'B']}$ are covariantly constant
tensors in $\wedge^2(E)$ and $\wedge^2(H)$ respectively,
which we use to raise and lower the $A,A'$ indices.
We will always choose local frames in $H$ such that $\eps_{12} = 1$ and the
Hermitean metric is $\eta_{A'{\bar B}'}=\delta_{A'}^{B'}$,
and denote the corresponding
coordinates in the fiber of $H$ by $\pi^{A'}$.  Similarly, our frames in $E$ will
always be orthonormal, $\eta_{A \bar{B}} = \delta_A^B$.

The spin connection 1-form on $\cM$ splits into
\be
\Omega_{AA';BB'} = \eps_{AB}\,p_{A'B'} + \eps_{A'B'}\,q_{AB}\,,
\ee
where $q_{AB} = q_{(AB)}$  and $p_{A'B'} = p_{(A'B')}$ are connection 1-forms valued respectively in ${\mathfrak{usp}}(2n) \subset S^2(E)$ and ${\mathfrak{usp}}(2) \subset S^2(H)$.
From the quaternionic vielbein one can construct the metric as well as
a triplet $\omega^{A'B'} = \omega^{(A'B')}$ of 2-forms:
\be
\de s^2_{\cM}= \eps_{AB}\,\eps_{A'B'}\,V^{AA'}\,V^{BB'}\,,\qquad
\omega^{A'B'} =  \frac{1}{2}\,\eps_{AB} \left( V^{AA'} \wedge V^{BB'} + V^{AB'} \wedge V^{BA'} \right)\,.
\ee
If $\cM$ were \hk, the $\omega^{A'B'}$ would be the \kahler forms for the three complex structures, and
in particular they would be separately closed.
In the quaternionic-\kahler case the triplet is covariantly closed with respect
to the $USp(2)$ connection:
\be
\de\omega^{A'}_{B'} + p^{A'}_{C'} \wedge \omega^{C'}_{B'} - p^{C'}_{B'} \wedge \omega^{A'}_{C'} = 0\,.
\ee
Moreover, the $USp(2)$ curvature is proportional to $\omega^{A'}_{B'}$:
\be \label{usp2-curvature}
\de p^{A'}_{B'} + p_{C'}^{A'} \wedge  p^{C'}_{B'} = \frac{\nu}{2} \omega^{A'}_{B'}\,.
\ee
Quaternionic-K\"ahler manifolds are always Einstein;
the constant $\nu$ appearing in \eqref{usp2-curvature} is related to the
scalar curvature
by $R=4n(n+2) \nu$, see \textit{e.g.} \cite{Bergshoeff:2002qk}.
We shall restrict to the negative
curvature case (this is always the case for the manifolds appearing
in sigma models coupled to ${\cal N}=2$ supergravity \cite{Bagger:1983tt}).

By contracting $\omega^{A'B'}$ with the metric one
obtains the quaternionic structure operators $J^{A'B'}$.
These satisfy the quaternionic algebra and are covariantly constant
with respect to the $USp(2)$ connection, but do not have
vanishing Nijenhuis tensor  (see \textit{e.g.} Appendix B in 
\cite{Bergshoeff:2002qk}).

There is a standard way to construct a \hk manifold of dimension $4n+4$, fibered over $\cM$:
namely, the total space $\cS$ of $H^\times / \Z_2$ over $\cM$, where $H^\times$ is
the $\C^2$ bundle $H$ with the zero section deleted, and $\Z_2$ acts as $\pi^{A'} \to - \pi^{A'}$
on the fiber of $H$.  $\cS$ is known as the ``Swann space'' or ``\hk cone'' of $\cM$ \cite{MR1096180,deWit:2001dj}.
Its metric is
\begin{equation}
\label{dshkc}
\de s^2_\cS = \abs{ \De \pi^{A'} }^2 + \frac{\nu}{4}\,r^2 \de s^2_\cM\,.
\end{equation}
Here, $r^2 \equiv \abs{\pi^1}^2 + \abs{\pi^2}^2$ is the $USp(2)$
invariant norm in the fiber of $H$, and
\be
\De \pi^{B'} \equiv \de \pi^{B'} + p^{B'}_{C'} \pi^{C'}
\ee
is the covariant differential of $\pi^{A'}$.
For $\nu < 0$, the case of interest in this paper, the  metric \eqref{dshkc}
has indefinite signature $(4,4n)$. In \cite{MR1096180} it is shown that
it is \hk, with \hk potential
(a simultaneous \kahler potential for all complex structures)\footnote{This
formula holds for quaternionic-K\"ahler spaces
with positive scalar curvature, but can easily be continued to
the case of negative scalar curvature by flipping the sign of
$r^2$ \cite{Bergshoeff:2005di}.}
\begin{equation} \label{hkpot-1}
\chi(x, \pi^{A'}, \bar{\pi}^{A'})= r^2\,.
\end{equation}
The metric \eqref{dshkc} admits a $SU(2)$ group of isometries, acting on the
$\IR^4 / \Z_2$ fiber by
\begin{equation}
\delta \pi^{A'} = \I \eps_3 \pi^{A'} + \eps_- \bar{\pi}^{A'}\,, \quad
\delta \bar{\pi}^{A'} = - \I \eps_3 \bar{\pi}^{A'} + \eps_+ \pi^{A'}\,,
\end{equation}
and a homothetic Killing vector $\pa_r$, equal to the
gradient of the \hk potential $\chi = r^2$.  It may also be written as
\begin{equation}
\de s^2_\cS = \abs{ \De \pi^{A'} }^2 + \frac{\nu}{2}\,\abs{\pi_{B'} V^{B B'}}^2\,
\end{equation}
reflecting one of the complex structures on $\cS$, for which
$\De \pi^{A'}$ $(A' \in \{1, 2\})$
and $\pi_{B'} V^{B B'}$ $(B \in
\{1, \dots, 2n\})$ are of type $(1,0)$ \cite{MR664330}, and the
\kahler form is
\begin{equation} \label{s-kahler-form}
\omega_\cS = \I\left( \De \pi^{A'} \wedge \De \bar \pi_{A'} + \frac{\nu}{2}\,\pi_{A'}\bar \pi_{B'} \omega^{A' B'} \right).
\end{equation}
In this paper we will always use this complex structure on $\cS$.
The other complex structures are obtained by rotating this one
under $SU(2)$; their respective K\"ahler forms can be obtained
by taking the real and imaginary part of the (2,0) form
\begin{equation}
\label{holsym}
\Omega = \De \pi^{A'} \wedge \De \pi_{A'} + \frac{\nu}{2}\,\pi_{A'} 
\pi_{B'} \omega^{A'B'}\,.
\end{equation}
$\Omega$ is not manifestly holomorphic, but indeed 
defines a holomorphic symplectic structure on $\cS$.  There is also a natural 
holomorphic Liouville form
\begin{equation} \label{can1-form}
\cX = \pi_{A'} \De \pi^{A'}\,,
\end{equation}
which obeys (using \eqref{usp2-curvature})
\begin{equation} \label{liouville-relations}
\de \cX = \Omega\,, \qquad \iota_\cE \Omega = 2 \cX\,,
\end{equation}
where $\cE$ is the ``Euler'' vector field
\begin{equation}
\cE = \pi^{A'} \frac{\pa}{\pa \pi^{A'}}\,.
\end{equation}

The cross-section of $\cS$ at a fixed value of the \hk potential
defines a 3-Sasakian space $\cal J$,
which is a $S^3$ fiber bundle
over the \qk space $\cM$. It is useful to view $S^3$ as a Hopf fibration
over $S^2$, and choose coordinates
\be
 \e^{\I\phi} =\sqrt{\pi^2/\bar \pi^2}\,,\quad
z=\pi^1/\pi^2\,,
\ee
on the $U(1)$ fiber and $S^2$ base respectively.
The metric \eqref{dshkc} can then be rewritten as
\be \label{dshkcr}
\de s^2_{\cS}=\de r^2 + r^2 \left[
\sigma_1^2+\sigma_2^2+ \sigma_3^2 + \frac{\nu}{4} \de s^2_{\cM} \right]
\ee
where $\sigma_i$ are a triplet of $1$-forms,
\be
\sigma_1+\I \sigma_2=\frac{\de z+{\cal P}}{1+\bar z z}
\,,\quad
\sigma_3 =  \de \phi - \frac{\I}{2(1+z\bar z)} \left(
\bar z \de z - z \de \bar z\right) - \frac{\I}{r^2} \pi^{A'} p^{B'}_{A'} \bar\pi_{B'} \,,
\ee
and ${\cal P}$ is the ``projectivized'' $USp(2)$ connection,
\be
{\cal P} =  p^1_2 + z(p^1_1 - p^2_2) - z^2 p^2_1\,.
\ee
In \eqref{dshkcr}, the term in brackets is the metric on $\cJ$.

By dividing out the $U(1)$ action on the fiber of ${\cal J}$ one obtains a
$(4n+2)$-dimensional space $\cZ = {\cal J} / U(1)$, the twistor space of $\cM$.
Since the $U(1)$ action preserves the K\"ahler form \eqref{s-kahler-form} on $\cS$ and
the complex structure on $\cS$ relates $\pa_\phi$ to the
homothetic Killing vector $\pa_r$, $\cZ$ can be
thought of as a K\"ahler quotient, $\cZ = \cS \sympquotient U(1)$.  
The \kahler metric is
\be
\label{dsz}
\de s^2_{\cZ} = \frac{|\de z +{\cal P}|^2}{(1+z\bar z)^2} + \frac{\nu}{4} \de s^2_{{\cal M}}\,,
\ee
with the K\"ahler form
\be
\omega_\cZ = 
\I \left( \frac{(\de z+{\cal P}) \wedge (\de \bar z+\bar {\cal P})}{(1+z\bar z)^2}
+ \frac{\nu}{2} \frac{\bar{\pi}_{A'} \pi_{B'}}{r^2} \omega^{A'B'}\right).
\ee
A K\"ahler potential $K_{\cZ}$ may be obtained from the \hk potential
$\chi$ on $\cS$ by writing
\begin{equation} \label{hk-potl}
\chi(\lambda,\bar \lambda,u,\bar u) = 
\abs{\lambda}^2  \e^{K_\cZ(u,\bar u)}\,,
\end{equation}
where $(u^i, \lambda)$ are complex coordinates on $\cS$, such that $u^i$
are inert under the $U(1)$ action and $\lambda$ transforms
with weight $1$.

As is well known, the \kahler quotient $\cZ = \cS \sympquotient U(1)$ may also be described as
$\cZ = \cS / \IC^\times$; the $\C^\times$ action is the one generated by $\cE$, $\pi^A \to \mu \pi^A$.
This realizes $\cZ$ as a complex manifold equipped with a natural
holomorphic line bundle, namely $\cS$ itself:  we call this line bundle $\cO(-2)$, and
its $k$-th power $\cO(-2k)$.\footnote{This notation is justified by the fact that $\cO(-2)$ restricts to
the usual line bundle $\cO(-2)$ over each $\C\PP^1$ fiber of $\cZ$.  Locally one can define a bundle
$\cO(-1)$ as well, namely the total space of $H^\times$ instead of $H^\times / \Z_2$.  However,
the decomposition $T_\C \cM = E \otimes H$ need not exist globally over $\cM$, so globally $\cO(-1)$
need not exist.}
A (holomorphic) function on $\cS$, homogeneous of degree $\ell$ under $\pi^A \to \mu \pi^A$, is thus
a (holomorphic) section of $\cO(\ell)$.
From this point of view, \eqref{hk-potl} is simply the statement 
$K_\cZ = \log \norm{s}^2$ where $s$ is a local holomorphic section of $\cO(-1)$
and $\norm{\cdot}$ is the norm induced from the one in $H$.

The Liouville form $\cX$ on $\cS$ descends to a complex contact structure on $\cZ$ \cite{MR664330}, 
given by an $\cO(2)$-valued holomorphic 1-form $X$.  
(Indeed, by definition an $\cO(2)$-valued 1-form on $\cZ$
is the same as a 1-form on $\cS$ which is homogeneous of degree $2$ in the $\pi^{A'}$ and
has zero inner product with $\cE$.)  Using \eqref{liouville-relations}, $X$ may also obtained by contracting 
$\Omega$ with $\half \cE$.  To represent it as 
a 1-form on $\cZ$ we must choose some local section of $\cO(2)$, thus locally trivializing the
line bundle.  One natural choice is given 
by the degree 2 homogeneous (non-holomorphic) function $\pi^1 \pi^2$.  Then a short computation
from \eqref{can1-form} gives
\be
\label{contact-form}
X = \frac{\de z + \mathcal{P}}{z}\,.
\ee
Dually, $\cZ$ has a holomorphic $\cO(-2)$-valued vector field,
the ``Reeb vector'' $Y$, determined by the conditions
\begin{equation} \label{reeb}
\iota_Y \de X = 0\,, \quad \iota_Y X = 1\,.
\end{equation}

Finally, one may obtain the \qk manifold $\cM$ by projecting the
metric \eqref{dsz} down to the base, orthogonally to $Y$ and its
complex conjugate. The process of going from $\cS$ to $\cM$ is known
in supergravity as the superconformal quotient. An important fact is
that isometries of $\cM$ compatible with the quaternionic structure lift to holomorphic isometries
of $\cZ$, and to tri-holomorphic isometries of $\cS$. More details on
the map between \hk and quaternionic-K\"ahler spaces
can be found in \cite{deWit:1999fp,deWit:2001dj,Bergshoeff:2004nf}.

\subsection{Tri-holomorphic isometries and projective superspace \label{3holg}}

In the last subsection we described how to start from a $4n$-dimensional quaternionic-\kahler
space $\cM$ and build up its $(4n+4)$-dimensional \hk cone $\cS$.
Here we consider the special case where $\cS$ admits $n+1$ commuting triholomorphic isometries.  In this case $\cS$ admits a very simple description, which from the physical
point of view comes from the duality between hypermultiplets and tensor multiplets in
four dimensions, and the existence of the (off-shell) projective
superspace formalism for tensor multiplets.  We review that description here;
in Section \ref{sec-covariant-cmap} we will use it to get geometric information about $\cS$ and $\cZ$.

So suppose $\cS$ is a \hk manifold of dimension $4(n+1)$ with $n+1$ commuting triholomorphic isometries. Then the metric is of the ``generalized
Gibbons-Hawking'' form \cite{Karlhede:1984vr,Hitchin:1986ea},
\be
\label{ds3hol}
\begin{split}
\de s^2_{\cS} =& {\cal L}_{x^I x^J} \left(\frac{1}{4}\de x^I \de x^J
+ \de v^I \de{\bar v}^J \right)\\
&+ \frac14 {\cal L}^{x^I x^J}
\left( \de\theta_I + \I {\cal L}_{v^K x^I}\de v^K
- \I {\cal L}_{x^I \bar v^K} \de \bar v^K \right)
\left( \de\theta_J + \I {\cal L}_{v^L x^I}\de v^L
- \I {\cal L}_{x^I \bar v^L} \de \bar v^L \right)\,.
\end{split}
\ee
In \eqref{ds3hol} we use coordinates $(v^I, \bar v^I, x^I, \theta_I)$ on $\cS$:
$v^I$ is complex and $x^I, \theta_I$ are real.  ${\cal L}$ is a function of
$(v^I,\bar v^I,x^I)$,
known as the ``tensor Lagrangian'' because of the way it enters the tensor multiplet formalism.
We also denoted ${\cal L}_{x^I x^J} \equiv \pa_{x^I}\pa_{x^J} {\cal L}$ etc,
and use ${\cal L}^{x^I x^J}$ for the inverse matrix to ${\cal L}_{x^I x^J}$.

The requirement that \eqref{ds3hol} is \hk, and moreover that it has
a homothetic Killing vector, leads to constraints
on $\cal L$ \cite{deWit:2001dj}:
${\cal L}$ must be homogeneous of degree $1$
in $(v^I,\bar v^I, x^I)$, and invariant under a common
phase rotation $v^I\to
 \e^{\I\vartheta} v^I$.  Furthermore, ${\cal L}$ must satisfy a set of linear
partial differential equations, given as Eqs. (5.10) in \cite{deWit:2001dj}.
Any solution of these constraints may be expressed as a contour integral
\begin{equation}\label{SS-lagr}
{\cal L}(v^I,\bar v^I,x^I)= {\rm Im} \oint_{\cC} \frac{\de
\zeta} {2 \pi \I \zeta}\; G(\eta^I(\zeta))\,,
\end{equation}
where $\eta^I$ are ``real $\cO(2)$ projective superfields'', written
\begin{equation}\label{eta}
\eta^I = \frac{v^I}{\zeta} + x^I - {\bar v}^I \zeta\,,
\end{equation}
and $G(\eta^I)$ is a holomorphic function, homogeneous of degree $1$
in its arguments, which we call the ``generalized prepotential''.  The
function $G$ and contour $\cC$ 
completely specify the local \hk geometry of $\cS$.

$\zeta$ can be thought of as a stereographic coordinate on the $\C\PP^1$
fiber of the twistor space over ${\cS}$ \cite{Hitchin:1986ea}.
As we shall see, at least for the cases we study in the next section, after
evaluating the contour integral by residues,
$\zeta$ becomes identified up to a phase with a natural
stereographic coordinate $z$
on the twistor space $\cZ$ over $\cM$.

To make one of the complex structures of $\cS$ explicit, one can trade
the real coordinates $(x^I, \theta_I)$ for the complex coordinates
\be
w_I = \frac12 \left( {\cal L}_{x^I} + \I \theta_I \right)\,.
\ee
Then the tri-holomorphic isometries $\theta_I\to \theta_I+\eps_I$
correspond to imaginary shifts of $w_I$.
The metric \eqref{ds3hol} is \hk, and its \hk potential
is the Legendre transform of ${\cal L}$ with respect to all $x^I$,
obtained by first computing
\begin{equation}\label{legendre-tr}
\chi (v^I,\bar v^I,\chi_I) \equiv {\cal L} (v,\bar v,x)  - \chi_I\,x^I, \qquad \chi_I
= \frac{\partial {\cal L}}{\partial x^I}\,.
\end{equation}
and then substituting $\chi_I = w_I + \bar w_I$ to obtain $\chi(v^I, w_I, \bar v^I, \bar w_I)$.
The \hk cone corresponds to an open domain in the
space $\IR^{4n+4}$ spanned by the variables $v^I,w_I$, bounded by
the tip of the cone $\chi=0$. 

$\chi$ is a function of $(v^I, \bar v^I, w_I+\bar w_I)$ 
only, and has scaling weight $2$ under the $\R^\times$ action
\begin{equation}
v^I \to \mu^2 v^I\,, \quad w_I \to w_I\,.
\end{equation}
Moreover, it is invariant under $SU(2)$ transformations acting
on $(v^I, w_I)$ as
\cite{deWit:2001dj}
\begin{equation}
\label{su2vw}
\delta v^I=\I\epsilon^3v^I+\epsilon^+x^I\,,\quad
\delta {\bar v}^I=-\I\epsilon^3{\bar v}^I+\epsilon^-x^I\,,\quad
\delta w_I= \epsilon^+ {\cal L}_{v^I}\,,\quad
\delta {\bar w}_I= \epsilon^- {\cal L}_{\bar v^I}\,,
\end{equation}
where $x^I$ is related to $(v^I,w_I,\bar v^I,\bar w_I)$
by the inverse Legendre transform,
\be
x^I = \frac{\pa\chi}{ \pa \chi_I}\,.
\ee
These transformations reflect the fact that
${\vec r}^I=(r^3,r^+,r^-)^I=(x^I,2v^I,2{\bar v}^I)$ transforms
linearly as a three-vector,
\begin{gather}
\label{su2xe}
\delta x^I=-2(\epsilon^-v^I+\epsilon^+{\bar v}^I)\,,\quad
\delta v^I=\I\epsilon^3v^I+\epsilon^+x^I\,, \quad \delta {\bar v}^I=-\I\epsilon^3{\bar v}^I+\epsilon^-x^I\,.
\end{gather}
The holomorphic symplectic and Liouville forms on $\cS$ take
the simple form \cite{deWit:2001dj}
\be
\Omega = {\rm d}w_I \wedge {\rm d}v^I\,,\quad \cX = v^I\,{\rm d}w_I\,,
\ee
so $(v^I, w_I)$ can be thought of as holomorphic Darboux coordinates for $\cS$.

As described in Section \ref{twiswa}, the twistor space $\cZ$
can be obtained by a $\C^\times$ quotient of $\cS$.  In the coordinates
$(v^I, w_I)$ the relevant $\C^\times$ action is
just complex multiplication on all the $v^I$ \cite{deWit:2001dj}.
So we can single out one coordinate, say $v^\flat$, and
define coordinates $(\xi^\Lambda, \txi_\Lambda, \alpha)$
(collectively denoted as $u_i$ in Section \ref{twiswa}) 
on $\cZ$ by
\be
\label{hkctoz}
v^\Lambda = v^\flat \xi^\Lambda\,,\quad
w_\Lambda = \frac{\I}{2} \txi_\Lambda\,,\quad
w_\flat  = \frac{1}{4\I} \left( \alpha + \xi^\Lambda \txi_\Lambda \right)\,.
\ee
In addition, $\lambda^2 = v^\flat$ defines
a local trivialization of $\cO(-2)$ over $\cZ$.
By homogeneity, the \hk 
potential $\chi$ factorizes as in \eqref{hk-potl},
\begin{equation}\label{chi-with-isom}
\chi = \sqrt{v^\flat {\bar v}^\flat}\,  \e^{K_{\cal Z}(u, \bar u)}\, .
\end{equation}
We will use this relation in Section \ref{quartic} to determine $K_\cZ$.

Finally, expressing the holomorphic symplectic form $\Omega$ in
terms of $v^\flat$ and $(\xi^\Lambda, \txi_\Lambda, \alpha)$, and
contracting with $\pa_{v^\flat}$, gives the holomorphic contact form on $\cZ$
and its associated Reeb vector \cite{deWit:2001dj},
\be \label{contact-form-local}
X=2(\de\alpha + \txi^\Lambda \de\xi_\Lambda - \xi^\Lambda \de\txi_\Lambda)\, ,
\quad Y = \half \frac{\pa}{\pa \alpha}\,.
\ee
These expressions are valid in the holomorphic trivialization $v^\flat=1$,
and motivated the introduction of the coordinates in \eqref{hkctoz}.

\subsection{$c$-map spaces and their generalized prepotentials\label{cmapg}}

We now specialize to the case where the quaternionic-K\"ahler manifold
$\cM$ arises by applying the $c$-map to a projective (\textit{i.e.} non-rigid) special K\"ahler manifold $\cM_s$.

First recall that locally the geometry of $\cM_s$ is determined by a single holomorphic function $F(X^\Lambda)$, homogeneous of degree two ($\Lambda \in \{0, \dots, n-1\}$).  Namely, choosing a symplectic section
$(X^\Lambda(z), F_\Lambda(X(z)) \equiv \pa F/\pa X^\Lambda)$ over $\cM_s$, the metric in $\cM_s$ is
\begin{subequations}
\begin{gather}
\cG_{a {\bar b}} =
\partial_{a} \partial_{{\bar b}} \cK(X(z), \bar X(\bar z))\,, \\
\cK (X, \bar X) \equiv -\log K(X, \bar X)\,,\\
K(X, \bar X) \equiv X^\Lambda N_{\Lambda\Sigma} \bar X^\Sigma\,,
\end{gather}
\end{subequations}
where $N_{\Lambda\Sigma}(X,\bar X)$ is given by the usual
special geometry formula
\begin{equation} \label{def-N}
N_{\Lambda\Sigma} \equiv \I \left(F_{\Lambda\Sigma} -
\bar{F}_{\Lambda\Sigma} \right),\qquad
F_{\Lambda\Sigma}\equiv \pa_{X^\Lambda}\pa_{X^\Sigma} F\,.
\end{equation}

Now we define $\cM$ locally as an $\IR^{2n+1}$ bundle
over $\IR^\times \times \cM_s$:
the fiber is parameterized by $2n+1$ real coordinates
$(\zeta^\Lambda, \tzeta_\Lambda, \sigma)$,
the $\R^\times$ factor by a real coordinate $ \e^U$.
The \qk metric on $\cM$ is then \cite{Cecotti:1988qn,Ferrara:1989ik}
\begin{eqnarray}\label{QK-metric}
\de s^2_\cM &=& 4\,\de U^2 - {\e}^{-2 U}({\cal N}+{\bar {\cal
N}})_{\Lambda\Sigma} W^\Lambda{\bar W}^\Sigma+\frac{1}{16}\,{\e}^{-4 U}\Big(
\de \sigma +\tzeta_\Lambda\de\zeta^\Lambda
- \zeta^\Lambda\de\tzeta_\Lambda \Big)^2
\nonumber\\
&& +\,4\,{\cal G}_{a\bar b}\,\de z^a \de{\bar z}^{\bar b}
\end{eqnarray}
where we defined
\begin{equation}\label{def-curlyN}
{\mathcal N}_{\Lambda\Sigma} \equiv - \I \bar
F_{\Lambda\Sigma} - \frac{(N X)_\Lambda (N X )_\Sigma}{(X N X)}\,, \quad
W^\Lambda \equiv ({\cal N}+{\bar {\cal N}})^{-1\,\Lambda\Sigma}
\Big({\bar{\cal N}}_{\Sigma\Pi} \de\zeta^\Pi
+ \I \de \tzeta_\Sigma\Big).
\end{equation}
This space admits isometries acting on the $\IR^{2n+1}$ fiber, 
\be
\label{heisqk}
P^\Lambda = \pa_{\tzeta_\Lambda} - \zeta^\Lambda \pa_\sigma\,,\quad
Q_\Lambda = -\pa_{\zeta^\Lambda} - \tzeta_\Lambda \pa_\sigma\,,\quad
K = \pa_\sigma\,,
\ee
which satisfy the Heisenberg algebra
\be
\label{heisalg}
[P^\Lambda, Q_\Sigma] = -2\,\delta^\Lambda_\Sigma\, K\,.
\ee
In addition, it has a non-compact isometry
\be
H=-\pa_U - \zeta^\Lambda \pa_{\zeta^\Lambda} - 
\tzeta_\Lambda \pa_{\tzeta_\Lambda} - 2 \sigma \pa_{\sigma}
\nn
\ee
which grades the Heisenberg algebra, 
\be
[H,P^\Lambda]= P^\Lambda, \quad [H,Q_\Lambda]=Q_\Lambda\ ,\quad
[H,K]=2 K
\nn
\ee

Physically, the metric \eqref{QK-metric} describes the classical
moduli space of the $D=3$ theory obtained by
beginning with a $D=4$, $\CN=2$ supergravity theory
coupled to $n-1$ Abelian vector multiplets, then reducing the theory
along a spacelike direction. In this context, the scalars $z^a$
are the moduli of the $D=4$
theory (\textit{e.g.} \kahler or complex structure
moduli for Type IIA or IIB respectively compactified on a Calabi-Yau
threefold),
$ \e^U$ is the radius of the fourth direction, $\zeta^\Lambda$
are the fourth component of the gauge fields, $\tzeta_\Lambda$
are the Poincar\'e duals of the $D=3$ one-forms coming from
the reduction of the vector fields in 4 dimensions, and $\sigma$ is the
dual of the Kaluza-Klein connection. Worldsheet instantons in general
break the isometries \eqref{heisqk}.

The same type of metric also occurs as the tree-level hypermultiplet moduli
space already in 4 dimensions, with a different interpretation of
the coordinates:  now
$z^a$ are complex structure or \kahler moduli respectively in Type IIA or IIB,
$ \e^{2U}= \e^{\phi}$ is the four-dimensional dilaton, and $(\zeta^\Lambda, \tzeta_\Lambda, \sigma)$
are scalars coming from the Ramond-Ramond sector. Space-time
instantons are now responsible for the breaking of the
isometries \eqref{heisqk}. T-duality along the fourth dimension exchanges
these two descriptions of the moduli space.

Finally, the analytic continuation $(\zeta^\Lambda,
\tzeta_\Lambda) \to \I (\zeta^\Lambda, \tzeta_\Lambda)$ of the
metric \eqref{QK-metric}, which arises from the reduction of
$D=4$, $\CN=2$ supergravity coupled to $n-1$ Abelian vector 
multiplets along a
\textit{timelike} direction, is relevant to the study of stationary
 black hole solutions \cite{Gunaydin:2005mx}.
This pseudo-Riemannian manifold, dubbed the $c^*$-map of $\cM_s$ in 
\cite{Gunaydin:2005mx}, is of a type called ``para-\qk'' in the mathematical 
literature (see \textit{e.g.} \cite{MR2143244} for a recent review, and 
\cite{Cortes:2005uq} for
an extensive discussion of the rigid $c^*$-map).

Having defined $\cM$ we now turn to its description in the tensor multiplet formalism.
In \cite{Rocek:2005ij,Rocek:2006xb}
(see also \cite{Berkovits:1995cb,Berkovits:1998jh}), it was shown that the
quaternionic-K\"ahler metric \eqref{QK-metric} is determined by a
generalized prepotential
\begin{equation}\label{G-F}
G(\eta^I)=\frac{F(\eta^\Lambda)}{\eta^\flat}\,,
\end{equation}
where $F(\eta^\Lambda)$ is a prepotential for the special K\"ahler manifold
$\cM_s$. Here, the indices $\Lambda \in \{0, \dots, n-1\}$ as usual in special
geometry, while the indices $I$ take one extra value,
$I \in \{ \flat, 0, \dots, n-1 \}$. Physically, the projective
superfield $\eta^\flat$ describes the compensating tensor multiplet used
in superconformal calculus (see Appendix A).
Geometrically, it provides the extra
quaternionic variable which extends the quaternionic-K\"ahler $\cM$ to
its Swann space $\cS$.  In order to prove \eqref{G-F},
the authors of \cite{Rocek:2005ij,Rocek:2006xb} evaluated the Legendre
transform
\eqref{legendre-tr} in a particular $SU(2)$ gauge, where
\be
\label{singgauge}
v^\flat = 0\,.
\ee
After performing the superconformal quotient, they found agreement with the
metric \eqref{QK-metric} upon identifying\footnote{Compared to
\cite{Rocek:2005ij,Rocek:2006xb}, we have set $\phi=2U, A^\Lambda=
\frac12 \zeta^\Lambda, B_\Lambda=-\tzeta_\Lambda$ and multiplied
$\sigma$ by $\qtr$\label{rvv-vs-us}.}
\begin{subequations}
\label{rvv}
\bea
X^\Lambda &=& \frac{v^\Lambda}{\sqrt{x^\flat}}\,,\quad
 \e^{2U}=\frac{K(X,\bar X)}{4 x^\flat}\,,\\
\zeta^\Lambda &=& \frac{x^\Lambda}{x^\flat}\,,\quad
\tzeta_\Lambda =  -\I (w_\Lambda - \bar w_\Lambda ) - \half
(F_{\Lambda\Sigma}+\bar F_{\Lambda\Sigma}) \zeta^\Sigma\,,\\
\sigma &=& 2 \I ( w_\flat-\bar w_\flat ) + \zeta^\Lambda \tzeta_\Lambda
+ \frac12 \zeta^\Lambda (F_{\Lambda\Sigma}+\bar F_{\Lambda\Sigma}) \zeta^\Sigma\,.
\eea
\end{subequations}
Moreover, the hyperk\"ahler potential
in the limit \eqref{singgauge} was found to be
\be
\label{chihkclim}
\chi(v^I,\bar v^I,w_I,\bar w_I) = \sqrt{2\ K(v^\Lambda,\bar v^\Lambda)
\ \left[ (w+\bar w)_{\Lambda} N^{\Lambda\Sigma} (w+\bar w)_{\Sigma}
- (w+\bar w)_\flat \right]}\,,
\ee
and could be rewritten in a much simpler way as
\be
\label{hksing}
\chi(v,\bar v,w,\bar w) = K\left[ X^\Lambda(v,\bar v,w,\bar w),
\bar X^\Lambda(v,\bar v,w,\bar w) \right].
\ee
It should be stressed that the gauge-fixing \eqref{singgauge} is only suitable
for the purpose of evaluating the metric on the base:
it cannot be used directly to obtain the metric on the hyperk\"ahler cone
or on the twistor space.
In the next section, we repeat the analysis of
\cite{Rocek:2005ij}, without making the gauge choice \eqref{singgauge}.

\section{K\"ahler potentials, covariant $c$-map and twistor 
map} \label{sec-covariant-cmap}
In this section, we apply the recipe outlined in Section \ref{3holg}
to the generalized prepotential \eqref{G-F}.
In Sections 3.1, 3.2 and 3.3,
we evaluate the contour integral \eqref{SS-lagr},
take its Legendre transform, and  obtain the hyperk\"ahler
potential $\chi$ on the hyperk\"ahler cone ${\cS}$,
as well as the K\"ahler potential $K_\cZ$ on the twistor space $\cZ$.
In Section 3.4, we perform the superconformal quotient from 
$\cS$ to $\cM$, and identify the real coordinates on $\cM$ as $\IR^\times 
\times SU(2)$ functions on $\cS$; we refer to the relations \eqref{covcmap}
as the ``covariant $c$-map''.
Some of the technical
details of the derivation are presented in Appendix A.
In Section 3.5, we work out the converse, and express the 
complex coordinates on $\cS$ and $\cZ$ in terms of the 
real coordinates on $\cM \times \IR^4$ and $\cM \times S^2$,
respectively; we refer to the relations \eqref{gentwi}
as the ``twistor map''.

\subsection{The tensor Lagrangian}
We start by evaluating the tensor Lagrangian \eqref{SS-lagr} based
on the generalized prepotential \eqref{G-F}, without making any gauge
choice. The tensor Lagrangian is the imaginary part of
\begin{equation}
\label{contint}
{\cal I}=\oint_{\cC_+} \frac{\de \zeta}{2\pi \I \zeta}
\frac{F(\eta^\Lambda)}{\eta^\flat}= \oint_{\cC_+} \frac{\de \zeta}{2\pi
\I \zeta^2} \frac{F(\zeta \eta^\Lambda)} {\zeta\eta^\flat}\,,
\end{equation}
where the contour $\cC_+$ is taken to be a small circle around a root $\zeta_+$ of $\zeta \eta^\flat$,
given in \eqref{eta} as
\begin{equation}
\zeta \eta^\flat= -\bar v^\flat(\zeta-\zeta_+)(\zeta-\zeta_-)
\end{equation}
where
\begin{equation}
\zeta_{\pm}=\frac{x^\flat\mp r^\flat}{2{\bar v}^\flat}\,,\qquad
\zeta_+-\zeta_- =-\frac{r^\flat}{{\bar v}^\flat}\,,\qquad
\zeta_+\zeta_-=-\frac{v^\flat}{{\bar v}^\flat}\,,
\end{equation}
with $r^\flat=\sqrt{(x^\flat)^2+4v^\flat {\bar v}^\flat}$. Note in particular that
the two roots are antipodes on $\C\PP^1$:
\be
\label{zpmr0}
\zeta_+ \overline{\zeta_-}=-1\,.
\ee
It will be convenient also to introduce the real quantities
\be
\label{ctc}
C = r^\flat - x^\flat \,,\quad \tilde C=r^\flat + x^\flat.
\ee
We can now easily do the contour integral \eqref{contint} and find
\begin{equation}
\label{iint}
{\cal I}= \frac{F(\eta^\Lambda(\zeta_+))}{r^\flat}\,,
\end{equation}
with
\begin{equation}
\label{auseeta}
\eta^\Lambda(\zeta_{\pm})=
x^\Lambda-\frac{x^\flat}{2}\Big(\frac{v^\Lambda}{v^\flat}
+\frac{{\bar v}^\Lambda}{{\bar v}^\flat}\Big)\pm \frac{r^\flat}{2}
\Big(-\frac{v^\Lambda}{v^\flat}
+\frac{{\bar v}^\Lambda}{{\bar v}^\flat}\Big)\,,
\end{equation}
which obeys
\begin{equation}
\overline{[\eta^\Lambda(\zeta_{+})]}=\eta^\Lambda(\zeta_{-})\,.
\end{equation}
Taking the imaginary part of \eqref{iint}, we find the tensor Lagrangian
\begin{equation}\label{cont-cmap}
\framebox{$
{\cal L}(v,\bar v,x)=-\frac{\I}{2r^\flat}\Big(F(\eta^\Lambda_+)-
\bar F(\eta^\Lambda_-)\Big) $}
\end{equation}
Here and henceforth, we denote
$\eta^\Lambda_{\pm} \equiv \eta^\Lambda(\zeta_{\pm})$.

To compute the \hk potential $\chi$
we have to take the Legendre transform of \eqref{cont-cmap}
with respect to $x^I$. Using the homogeneity of $F$, which gives
\begin{equation}
2F(\eta_+)=\eta^\Lambda_+ F_\Lambda(\eta_+)\,,\qquad
F_{\Lambda}(\eta_+)=\eta^\Sigma_+ F_{\Lambda\Sigma}(\eta_+)\,,
\end{equation}
it is an easy exercise to show that (with $K(\eta_+,\eta_-) \equiv
\eta_-^\Lambda N_{\Lambda\Sigma} \eta_+^\Sigma$)
\begin{align}\label{osvtrick}
{\cal L}(v,\bar v,x) &= \frac{1}{4r^\flat}K(\eta_+,\eta_-)+\frac{1}{2}(\eta^\Lambda_++\eta^\Lambda_-)
\pa_{x^\Lambda}{\cal L} \\
&= x^\Lambda \pa_{x^\Lambda}{\cL} + \frac{1}{4r^\flat}K(\eta_+,\eta_-) - \frac{x^\flat}{2}\left(\frac{v^\Lambda}{v^\flat} + \frac{\bar{v}^\Lambda}{\bar{v}^\flat}\right) \pa_{x^\Lambda}{\cL} \label{osvtrick-2}\,.
\end{align}
From \eqref{osvtrick-2} it follows directly that the Legendre transform of $\cL$ with respect to
$x^\Lambda$ is
\begin{equation}
\label{1stleg}
\ltr{\cL - x^\Lambda \chi_\Lambda}_{x^\Lambda} =
\frac{1}{4 r^\flat} K(\eta_+,\eta_-) -\frac{x^\flat}{2} \left( \frac{v^\Lambda}{v^\flat}
+ \frac{\bar v^\Lambda}{\bar v^\flat}\right) \chi_{\Lambda}\,\bigg\vert_{\frac{\pa \cL}{\pa x^\Lambda} = \chi_\Lambda}\,,
\end{equation}
where we introduced the ``magnetic potential''\footnote{This terminology
anticipates the relation between $\chi_\Lambda$ and the magnetic charge
$p^\Lambda$, explained in Sections \ref{sec-leg} and \ref{bpsgeo}.}
$\chi_\Lambda$ as the conjugate
to $x^\Lambda$, and the angle brackets indicate that we
evaluate at a critical value of $x^\Lambda$.

To finish the Legendre transform computation of the \hk potential
$\chi$ from $\cL$, we would need to transform over
the remaining $x$ variable $x^\flat$.  In principle one could do this by
first expressing the $x^\Lambda$ as functions of $(v^I,\bar v^I,\chi_\Lambda,x^\flat)$,
substituting these expressions in $K(\eta_+,\eta_-)$, and then
directly computing the transform over $x^\flat$.
In the next subsection we will see a more elegant way forward.

\subsection{Legendre transform, Hesse potential and black hole entropy
\label{sec-leg}}

In the last section we introduced the magnetic potentials $\chi_\Lambda$, determined by extremizing
the left side of \eqref{1stleg} to be
\begin{equation}\label{chilambda}
\chi_\Lambda = \frac{\pa \cL}{\pa x^\Lambda} = -\frac{\I}{2r^\flat}\Big(F_\Lambda(\eta_+) - \bar F_\Lambda(\eta_-) \Big).
\end{equation}
It turns out to be convenient
(as suggested by symplectic invariance) to introduce 
as well the ``electric potentials''
\be \label{electric-potentials} \phi^\Lambda \equiv
-\frac{\I}{2r^\flat}\Big(\eta^\Lambda_+-{\eta}^\Lambda_-\Big)
= \frac{\I}{2} \left( \frac{v^\Lambda}{v^\flat} - \frac{\bar
v^\Lambda}{\bar v^\flat}\right),
\ee
so that $\tilde\phi^\Lambda \equiv r^\flat \phi^\Lambda$ and $\tilde\chi_\Lambda \equiv r^\flat \chi_\Lambda$ are related to $\eta^\Lambda_\pm$ by
\be
\label{stabphi}
\begin{pmatrix} \tilde\phi^\Lambda \\ \tilde\chi_\Lambda  \end{pmatrix}
= \Im
\begin{pmatrix} \eta_+^\Lambda \\ F_\Lambda(\eta_+) \end{pmatrix}.
\ee
This equation, which determines the complex variable $\eta_+^\Lambda$
(at least locally) in terms of the real quantities $(\tilde\phi^\Lambda,
\tilde\chi_\Lambda)$, is familiar from the study of the attractor
mechanism in $\CN=2$ supergravity \cite{Ferrara:1995ih,Ferrara:1996um,
Ferrara:1997tw,Ooguri:2004zv}; namely, in the 
geometry of a BPS black
hole with charges $(p^\Lambda, q_\Lambda)$,
the properties of the horizon are determined by solving the equations
\begin{equation} \label{attr-charges}
\begin{pmatrix} p^\Lambda \\ q_\Lambda  \end{pmatrix}
= \Re
\begin{pmatrix} X^\Lambda \\ F_\Lambda(X) \end{pmatrix}
\end{equation}
for $X^\Lambda$:  the moduli at the horizon are given by the ratios of
the $X^\Lambda$, while the tree-level Bekenstein-Hawking
entropy\footnote{We have chosen a convenient
normalization for Newton's constant, $G_N=\pi$\,.} of the black hole is
given by
\begin{equation} \label{entropy-charges}
K(X, \bar X) = 4\,\Sigma(p^\Lambda,q_\Lambda)\,,
\end{equation}
where $\Sigma(p^\Lambda,q_\Lambda)$ is a homogeneous function of
degree 2 of the charge vector $(p^\Lambda,q_\Lambda)$, invariant
under symplectic transformations. 
The converse of the map \eqref{attr-charges} is then given by
\be
\label{solstab}
X^\Lambda = p^\Lambda + \I \frac{\pa\Sigma(p,q)}{\pa q_\Lambda}\,,\quad
F_\Lambda(X) = q_\Lambda - \I \frac{\pa\Sigma(p,q)}{\pa p^\Lambda}\,.
\ee
The function $\Sigma(p^\Lambda,q_\Lambda)$
is also familiar as the Hesse potential of 
rigid special K\"ahler geometry, where it determines the metric 
in real coordinates \cite{Hitchin:2005uu,LopesCardoso:2006bg,Ferrara:2006js}.

Applying the above to \eqref{stabphi}, and making
use of the homogeneity of $F$, we find
\begin{equation} \label{ksbh}
K(\eta_+, \eta_-) = 4 (r^\flat)^2\, \Sigma(\phi^\Lambda, \chi_\Lambda)
\end{equation}
and
\be
\eta_+^\Lambda = r^\flat \left( \I \phi ^\Lambda
- \frac{\pa\Sigma(\phi,\chi)}{\pa \chi_\Lambda} \right),\quad
F_\Lambda(\eta_+) = r^\flat \left( \I \chi_\Lambda
+  \frac{\pa\Sigma(\phi,\chi)}{\pa \phi^\Lambda} \right).
\ee
In our computation of the Legendre transform below, the
form \eqref{ksbh} of $K$ will be useful,
because all the $x^\flat$ dependence has been isolated in the prefactor.

\subsection{Potentials on hyperk\"ahler cone and twistor space} \label{quartic}

Substituting the result \eqref{ksbh} in \eqref{1stleg},
we now consider the remaining Legendre transform with respect to $x^\flat$,
\be
\ltr{\label{legx0} {\cal L} - x^\Lambda \chi_\Lambda - x^\flat \chi_\flat}_{x^\Lambda, x^\flat} =
\ltr{r^\flat \Sigma(\phi^\Lambda, \chi_\Lambda) - x^\flat \tilde\chi_\flat}_{x^\flat}\,,
\ee
where we defined
\be \label{tilde-chi-flat} \tilde \chi_\flat  \equiv
 \chi_\flat  + \frac12
\left( \frac{v^\Lambda}{v^\flat} + \frac{\bar v^\Lambda}{\bar
v^\flat}\right) \chi_{\Lambda}\,.
\ee
We want to determine the value of $x^\flat$ extremizing the right side.
Noting that $\phi^\Lambda$ and $\chi_\Lambda$ are independent of $x^\flat$ (the latter by definition of the Legendre transform), we find
\be \label{solx0}
x^\flat
= \pm \frac{2 \sqrt{v^\flat \bar v^\flat}\,\tilde\chi_\flat }
{\sqrt{\Sigma^2(\phi,\chi)-\tilde \chi_\flat ^2}}\,,
\ee
We assume that the term under the square root is positive definite in the
region of interest, and choose the upper sign below; while a solution with
the opposite sign does in principle exist, it leads to a much more complicated
form of the \hk potential. 
From \eqref{solx0} it easily follows that
\be r^\flat= \frac{2 \sqrt{v^\flat \bar v^\flat}
\,\Sigma (\phi,\chi)} {\sqrt{\Sigma^2(\phi,\chi) - \tilde\chi_\flat ^2}}\,.
\ee
We then find that
\eqref{legx0} simplifies to
\be \label{chi-first}
\ltr{ {\cal L} - x^\Lambda \chi_\Lambda - x^\flat \chi_\flat }_{x^\Lambda, x^\flat} = 2 \sqrt{v^\flat \bar v^\flat
\left( \Sigma^2(\phi,\chi) -\tilde \chi_\flat ^2\right)}\,.
\ee
As described below \eqref{legendre-tr}, the hyperk\"ahler potential $\chi$ 
is obtained from this Legendre
transform upon replacing $\chi_I$ by $w_I + {\bar w_I}$.
Inserting as well the definitions \eqref{electric-potentials}, \eqref{tilde-chi-flat} 
of $\phi^\Lambda$ and $\tilde\chi_\flat$, 
we conclude that $\chi$ is given in terms of the complex variables $(v^I, w_I)$ by
\be
\label{chihkc}
\framebox{$
\begin{array}{cl}
\chi(v,\bar v,w,\bar w) = 2 \sqrt{v^\flat \bar v^\flat}  & \left\{ 
\Sigma^2\left[ \frac{\I}{2}
\left( \frac{v^\Lambda}{v^\flat} - \frac{\bar v^\Lambda}{\bar
v^\flat}\right), w_\Lambda+{\bar w}_\Lambda
\right] \right. \\
& \left. - \left[w_\flat +{\bar w}_\flat + \frac{1}{2} \left(
\frac{v^\Lambda}{v^\flat} + \frac{\bar v^\Lambda}{\bar v^\flat}\right)
(w_\Lambda+{\bar w}_\Lambda) \right]^2 \right\}^{\frac12}
\end{array}
$}
\ee
The condition that the quantity in curly brackets be strictly positive 
defines an open set in $\IR^{4n+4}$, which we identify as
the Swann space of $\cM$.

Thus, the \hk potential of the Swann space $\cS$ is simply expressed
in terms of the Hesse potential $\Sigma$, or equivalently
the Bekenstein-Hawking entropy functional.\footnote{The relation of $\chi$ 
to the black hole entropy will find a natural physical explanation in 
Section \ref{bpsgeo}, Equation \eqref{bhesse}.}
This parallels the rigid case, 
where the Hesse potential $\Sigma$ 
is known to provide a  K\"ahler potential for the
rigid $c$-map in complex Darboux 
coordinates \cite{Cecotti:1988qn,Hitchin:2005uu,Ferrara:2006js}.

We note that the \hk potential \eqref{chihkc}, and therefore the
metric on $\cS$, are invariant under the Killing vector fields
\be
\label{heishkc}
P^\Lambda = \frac{\I}{2}\,  \pa_{w_\Lambda}+ \cc\ ,\quad
Q_\Lambda =  -v^\flat \pa_{v^\Lambda} + w_\Lambda \pa_{w_\flat} + \cc\ ,
\quad K = -\frac{\I}{4}\, \pa_{w_\flat} + \cc
\ee
satisfying the Heisenberg algebra \eqref{heisalg}. Here, $P^\Lambda$
and $K$ are the $n+1$ triholomorphic isometries afforded by the
tensor multiplet description, while $Q_\Lambda$ are additional 
triholomorphic isometries acting as~\cite{Robles-Llana:2006ez,Berkovits:1995cb}
\begin{equation}
\label{tensiso}
\eta^\Lambda\rightarrow \eta^\Lambda + \epsilon^\Lambda \eta^\flat\,.
\end{equation}
Indeed, under this shift, the generalized prepotential \eqref{G-F}
changes by a term $\epsilon^\Lambda F_\Lambda(\eta)$ regular 
at $\zeta=\zeta_+$.
In Section \ref{sec-covariant-cmap}
we show that \eqref{heishkc} descend to the isometries \eqref{heisqk} 
of $\cM$ under the superconformal quotient.

We may obtain another useful 
expression for $\chi$ by switching back from $w^\flat$ to the
tensor multiplet variable $x^\flat$.  Namely, using \eqref{solx0} and \eqref{chi-first} gives directly
\begin{equation}
\label{chi-second}
\chi = 4 \frac{v^\flat \bar{v}^\flat}{r^\flat} \Sigma(\phi,\chi)\,.
\end{equation}
Using the relation \eqref{ksbh} between $\Sigma$
and the \kahler potential on $\cM_s$, this can also be written
\begin{equation}
\label{chiK}
\chi = \frac{v^\flat \bar{v}^\flat}{(r^\flat)^3} K\Big(\eta_+^\Lambda, \eta_-^\Lambda \Big).
\end{equation}
This relation between the geometry of $\cS$ and that of the
special \kahler manifold $\cM_s$ will become useful 
in Section \ref{sec-twistor-map}.

To discuss the twistor space $\cZ$, we use the coordinates
$(\xi^\Lambda, \txi_\Lambda, \alpha)$ introduced in
\eqref{hkctoz}. Plugging them into \eqref{chihkc}, one finds
the form of \eqref{chi-with-isom}, with
\be
\label{kz-hesse}
\framebox{$K_{{\cal Z}}= \frac12\log\left\{
 \Sigma^2 \left[\frac{\I}{2}(\xi^\Lambda-\bar\xi^\Lambda),
\frac{\I}{2}(\txi_\Lambda-\bar\txi_\Lambda)\right]
+\frac{1}{16} \left[
\alpha -\bar\alpha+ \xi^\Lambda \bar\txi_\Lambda-\bar\xi^\Lambda \txi_\Lambda
\right]^2 \right\}+\log 2$}
\ee
So, as for $\chi$, the K\"ahler potential on the twistor space $\cZ$ is simply expressed
in terms of the Hesse potential $\Sigma$ on the special K\"ahler 
manifold $\cM_s$ (or, rather, its rigidification $\cM_s'$). 
The range of $(\xi^\Lambda, \txi_\Lambda, \alpha)$ is restricted 
to the domain where
the sign of the bracket in \eqref{kz-hesse} is positive:  other 
values do not correspond to points of $\cZ$. The triholomorphic
isometries \eqref{heishkc} of $\cS$ descend to holomorphic 
isometries of $\cZ$, generated by the vector fields
\be
\label{heisz}
P^\Lambda = \pa_{\txi_\Lambda} - \xi^\Lambda \pa_{\alpha}+\cc\,,\quad
Q_\Lambda = -\pa_{\xi^\Lambda} - \txi_\Lambda \pa_{\alpha}+\cc\,,\quad
K = \pa_\alpha +\cc\,.
\ee
This standard form of the Heisenberg action was an additional motivation for 
the change of variable \eqref{hkctoz}.

At this stage, we note that when the special K\"ahler space ${\cal M}_s$
is a Hermitian symmetric space $G/K={\rm Conf}(J)/{\rm Struc}(J)\times U(1)$,
corresponding to the case where the prepotential $F$ is the cubic
norm of a Jordan algebra $J$, the Hesse potential $\Sigma$ becomes
equal to the square root of the quartic invariant
of the conformal group ${\rm Conf}(J)$. The term in bracket is then
recognized
as the ``quartic light-cone''\footnote{We are grateful to M.~G\"unaydin
for extensive discussions on these constructions.}
$\cN_4(\xi,\txi,\alpha;\bar\xi,\bar\txi,\bar\alpha)$
introduced in \cite{Gunaydin:2000xr}; in that work, it was shown that the
locus $\cN_4=0$ is invariant under the action of a group ${\rm QConf}(J)$
containing
${\rm Conf}(J)\times SU(2)$ as a subgroup; in fact, $\log \cN_4$ 
changes by K\"ahler
transformations under this action.  This implies that
the twistor space $\cZ$ carries a holomorphic, isometric
action of ${\rm QConf}(J)$, and that the quaternionic-K\"ahler base is itself
a symmetric space ${\rm QConf}(J)/{\rm Conf}(J)\times SU(2)$.
This fact is at the root of the construction of the 
quaternionic discrete series representations of 
${\rm QConf}(J)$ \cite{MR1421947}.
In a separate paper \cite{gnpw-to-appear-2},
the constructions of \cite{Gunaydin:2000xr} will be revisited in 
light of this observation.

\subsection{The covariant $c$-map \label{sec-covcmap}}

Our next task is to construct the projection from the \hk cone $\cS$ to the \qk base $\cM$:
we will express the coordinates $(U, z^a, \bar z^{\bar a}, \zeta^\Lambda, \tzeta_\Lambda, \sigma)$
on $\cM$ as $(\IR^\times \times SU(2))$-invariant functions on $\cS$.
We first list some possible candidates, and then argue that they are indeed equal to the
coordinates on $\cM$, as determined by the $c$-map metric \eqref{QK-metric}.
Further details of the derivation are given in the Appendix.

First we construct a candidate for $U$.
The hyperk\"ahler potential $\chi$ itself is $SU(2)$ invariant, but
has weight $2$ under $\R^\times$; it can be made invariant
by dividing out by $r^\flat$,
\be
\label{phichi}
 \e^{2U} \equiv
\frac{\chi}{4 r^\flat} = \frac{\Sigma^2(\phi,\chi)-\tilde \chi_\flat ^2}
{4\Sigma(\phi,\chi)}\,.
\ee

Next, recall from \eqref{su2xe} that ${\vec r}^I = (x^I, 2v^I, 2 \bar v^I)$
transforms as an $SU(2)$ vector, and has weight $2$ under
$\R^\times$.  Hence we can construct candidates for $\zeta^\Lambda$
by taking $SU(2)$ invariant dot products,
\be
\label{etaA}
\zeta^\Lambda \equiv \frac{1}{(r^\flat)^2}({\vec r}^\flat\cdot {\vec r}^\Lambda)
= \frac{1}{(r^\flat)^2}\left( x^\flat x^\Lambda
+ 2 \bar v^\flat v^\Lambda + 2 v^\flat \bar v^\Lambda \right) \,.
\ee
Using \eqref{auseeta}, this may also be written as
\be
\zeta^\Lambda =\frac12
\left( \frac{v^\Lambda}{v^\flat} +  \frac{\bar v^\Lambda}{\bar v^\flat}
\right)
+ \frac{x^\flat}{(r^\flat)^2} \Re[\eta_+^\Lambda] 
= \frac12 ( \xi^\Lambda + \bar\xi^\Lambda ) 
+ \frac{x^\flat}{(r^\flat)^2} \Re[\eta_+^\Lambda] \,.
\label{etaA2-2}
\ee

Next we construct the coordinates $z^\Lambda$ on $\cM_s$.
Using \eqref{su2xe}, one may check that under $SU(2)$ transformations one has
\begin{equation}
\delta[{\zeta_+\eta^\Lambda_+}]=
\big(\I\epsilon^3-2\epsilon^-\zeta_+\big)[\zeta_+
\eta^\Lambda_+]\,.
\end{equation}
Since this is just an overall $\Lambda$-independent rescaling, we can construct ratios which are $SU(2)$ and scale invariant:
\begin{equation}
\label{etaz}
z^a \equiv \frac{\eta^a_+}{\eta^0_+}
\,,\qquad
a=1, \dots, n-1.
\end{equation}

The remaining coordinates $\tzeta_\Lambda$ and $\sigma$ are trickier
to obtain. Symplectic covariance suggests considering the ``electric''
counterpart of \eqref{etaA2-2},
\begin{equation}
\tzeta_\Lambda \equiv
- \I (w_\Lambda - {\bar w}_\Lambda)
+ \frac{x^\flat}{(r^\flat)^2} {\rm Re}[F_\Lambda(\eta_+)]
=\frac12( \txi_\Lambda + \bar\txi_\Lambda)
+ \frac{x^\flat}{(r^\flat)^2} {\rm Re}[F_\Lambda(\eta_+)]\,,
\label{etaB2-2}
\end{equation}
whose $SU(2)$ invariance can indeed be checked by a somewhat
tedious computation. Finally, an even more tedious computation
shows that 
\bea
\label{etasig}
\sigma &\equiv& 2\I (w_\flat -{\bar w}_\flat )
+ \I \Big(\frac{v^\Lambda}
{v^\flat}w_\Lambda - \frac{{\bar v}^\Lambda}{{\bar v}^\flat}
{\bar w}_\Lambda\Big)
-\frac{x^\flat}{(r^\flat)^2} {\rm Re}\Big( \eta^\Lambda_+\,\tzeta_\Lambda
- F_\Lambda (\eta_+)\, \zeta^\Lambda\Big) \\
&=& \frac12 ( \alpha+\bar\alpha) 
-\frac{x^\flat}{(r^\flat)^2} {\rm Re}\Big( \eta^\Lambda_+\,\tzeta_\Lambda
- F_\Lambda (\eta_+)\, \zeta^\Lambda\Big)
\eea
is also invariant under $SU(2)$.

We claim that the functions $(U,z^a,\bar z^{\bar a},\zeta^\Lambda,
\tzeta_\Lambda,\sigma)$ on $\cS$ which we have just constructed give 
the projection of $\cS$ down to $\cM$. As a consistency check,
it is straightforward to check that with these identifications, 
the triholomorphic isometries \eqref{heishkc} on $\cS$ descend
to the isometries \eqref{heisqk} on $\cM$. 
A direct proof would involve
performing the superconformal quotient construction in our coordinates and checking that
the resulting metric
matches \eqref{QK-metric}.  In Appendix A we check this explicitly for the
components of the metric along $\de\tzeta_\Lambda$ and $\de\sigma$;
other components are fixed by supersymmetry.

Moreover, given that it was already verified in \cite{Rocek:2005ij}
that the superconformal quotient of $\cS$ yields \eqref{QK-metric},
and the $v^\flat \to 0$ limit of the coordinate functions were determined 
there as \eqref{rvv}, we only need to check that the $v^\flat \to 0$
limit of our coordinates agrees with \eqref{rvv}.
In this limit, the poles $\zeta_+$ and $\zeta_-$ approach $0$ and $\infty$
as $-v^\flat/x^\flat$ and $x^\flat/\bar v^\flat$, respectively, so that 
\be
\label{limeta}
\eta_+ ^\Lambda = -\frac{v^\Lambda}{v^\flat}x^\flat + x^\Lambda 
+ O(v^\flat)\,,
\ee
It is straightforward to check that 
the coordinates $(U, z^a, \bar z^{\bar a}, \zeta^\Lambda)$ defined in 
this section indeed agree with \eqref{rvv}, whereas a similar
check for $(\tzeta_\Lambda, \sigma)$ necessitates taking into account 
the next-to-subleading term in 
\eqref{limeta}, due to the appearance of certain singular
terms in the $v^\flat \to 0$ limit. Moreover, \eqref{chiK} reduces to
\eqref{hksing} in this limit.

We conclude that the coordinates $(U, z^a,\bar z^{\bar a}, \zeta^\Lambda, 
\tzeta_\Lambda, \sigma)$ on the base $\cM$ are given 
in terms of the complex coordinates $(v^I, w_I, \bar{v}^I, \bar{w}_I)$ 
on the hyperk\"ahler cone $\cS$ (or equivalently, the
complex variables $(\xi^\Lambda,\txi_\Lambda,\alpha,
\bar\xi^\Lambda,\bar\txi_\Lambda,\bar\alpha)$
on the twistor space $\cZ$):
\be
\label{covcmap}
\framebox{$
\begin{array}{c}
\e^{2U} = \chi / ( 4 r^\flat) \ \quad, \qquad
z^a = \eta^a_+/\eta^0_+\\
\zeta^\Lambda = 
\frac12\left( \xi^\Lambda + \bar\xi^\Lambda \right)
+ \frac{x^\flat}{(r^\flat)^2} \Re[\eta_+^\Lambda]\\
\tzeta_\Lambda =
\frac12\left( \txi_\Lambda + \bar\txi_\Lambda \right)
+ \frac{x^\flat}{(r^\flat)^2} {\rm Re}(F_\Lambda(\eta_+))
\\
\sigma =\frac12 \left( \alpha+\bar\alpha\right)
-\frac{x^\flat}{(r^\flat)^2}\Big({\rm Re}[\eta^\Lambda_+]\tzeta_\Lambda
- {\rm Re}[F_\Lambda (\eta_+)]\zeta^\Lambda\Big)
\end{array}
$}
\ee
We call these relations the covariant $c$-map formulae.

\subsection{The twistor map} \label{sec-twistor-map}

So far we have seen that $\cS$ defined by \eqref{G-F}
is fibered over $\cM$, and constructed the projection map
explicitly, but we have not given any information about the coordinates in the $\R^4 / \Z_2$ fiber.
In this section we will show that, given
a choice of symplectic section $(X^\Lambda, F_\Lambda(X))$
over the special \kahler space $\cM_s$, there is a canonically
defined coordinate $z$ in $\cZ$, holomorphic on each twistor fiber.
Moreover we give formulas relating $z$ and the coordinates in $\cM$ to the
complex coordinates $(\xi^\Lambda, \txi_\Lambda, \alpha)$ in $\cZ$.
We refer to this transformation as the ``twistor map''.
We then construct a corresponding coordinate system in $\cS$, with two complex fiber coordinates
$(\pi^1, \pi^2)$, such that $\pi^1 / \pi^2 = z$ and $\abs{\pi^1}^2 + \abs{\pi^2}^2 = \chi$.
For applications such as the Penrose transform these coordinates are very convenient, as we
will see in Section \ref{pentran}.

We start by expressing $\eta^\Lambda_\pm$
in terms of $\zeta^\Lambda$:  a straightforward computation 
using \eqref{auseeta} and \eqref{etaA2-2} gives
\be
\eta^\Lambda_\pm =
\frac{v^\Lambda}{\zeta_\pm} + x^\Lambda-\bar v^\Lambda \zeta_\pm
= \frac{v^\Lambda}{\zeta_\pm} +
\frac{(r^\flat)^2 \zeta^\Lambda - 2 \bar v^\flat v^\Lambda - 2
v^\flat \bar v^\Lambda}{x^\flat}
-\bar v^\Lambda \zeta_\pm\,.
\ee
This equation can be used
to express $\xi^\Lambda$ in terms of $\eta^\Lambda_\pm$:
\be
\label{veta}
\xi^\Lambda = \zeta^\Lambda + \frac{1}{(r^\flat)^2} \left(
\frac{v^\flat}{\zeta_+} \eta_+^\Lambda
- \bar v^\flat \zeta_+ \eta_-^\Lambda  \right).
\ee

Now, suppose we choose a symplectic section $(X^\Lambda(z^a),F_\Lambda
(z^a))$ over the
special K\"ahler manifold $\cM_s$. From \eqref{etaz} we know this
section is proportional to
$(\eta_+^\Lambda,F_\Lambda(\eta_+))$, so there exists $z \in \C^\times$ with
\be
\label{defz}
\frac{v^\flat}{(r^\flat)^2 \zeta_+} 
\begin{pmatrix}\eta_+^\Lambda\\ F_\Lambda(\eta_+)\end{pmatrix} 
=2
\I\, \e^{U+\frac12{\cal K}(X, \bar X)} z^{-1} 
\begin{pmatrix}X^\Lambda\\ F_\Lambda(X) \end{pmatrix}
\,.
\ee
The reason for choosing the complicated prefactors in \eqref{defz} will become clear below.
Using the definition \eqref{phichi} of $U$ and the relation \eqref{chiK},
we can rewrite \eqref{defz} as
\be
\label{etapX}
e^{\frac12\cK(\eta_+,\bar \eta_+)}
\begin{pmatrix}\eta_+^\Lambda\\ F_\Lambda(\eta_+)\end{pmatrix} 
= \I\, \frac{\zeta_+}{z}\,
\sqrt{\frac{\bar v^\flat}{v^\flat}} \,
e^{\frac12\cK(X,\bar X)}
\begin{pmatrix}X^\Lambda\\ F_\Lambda(X) \end{pmatrix}\,.
\ee
Then applying $K(\cdot, \bar{\cdot})$ to both sides shows that
the modulus of $z$ is equal to that of $\zeta_+$,
\be
\label{zzbar}
z \bar z = \zeta_+ \bar \zeta_+ =  \frac{C}{\tC}\,,
\ee
where $C$ and $\tC$ were defined in \eqref{ctc}. 
Substituting \eqref{defz} and its conjugate in \eqref{veta}, and using \eqref{zzbar},
now establishes our first ``twistor map'' relation:
\be
\label{eqva}
\xi^\Lambda =  \zeta^\Lambda + 2\I\, \e^{U + \frac12\cK(X,\bar X)}
\left( z \bar X^{\Lambda} + z^{-1} X^{\Lambda} \right).
\ee
This relation expresses the complex coordinates $\xi^\Lambda$ on $\cZ$
in terms of the coordinates ($X^\Lambda$, $\bar{X}^\Lambda$, $U$, $\zeta^\Lambda$) on the
base $\cM$ and a coordinate $z$ in the twistor fiber.  The rationale
for the choice of prefactors in \eqref{defz} is now clear: the modulus
was chosen such that the ratio between the last two terms in \eqref{veta}
has modulus $|z|^2$, while the choice of phase ensures that
$\xi^\Lambda$ depends holomorphically on $z$ when the base coordinates are fixed.
In other words, the fiber over every point on the base is rationally
embedded in $\cZ$, a key property of any twistor construction.
Changing the symplectic section on $\cM$ by
$X \to  \e^f X$ transforms $z$ by the phase $ \e^{\half (f - \bar f)}$.

To compute $\txi_\Lambda$, defined in \eqref{hkctoz}, we first 
write $\txi_\Lambda = 
\tzeta_\Lambda - (2\I w_\Lambda + \tzeta_\Lambda)$ and
use the relation \eqref{etaB2-2} in the last term.  Using $w_\Lambda
+{\bar w}_\Lambda=\chi_\Lambda$ in \eqref{chilambda}, we then obtain
\be
\label{eqwb0}
\txi_\Lambda = \tzeta_\Lambda +
\frac{1}{(r^\flat)^2}\left(
\frac{v^\flat}{\zeta_+} F_\Lambda(\eta_+) - \bar v^\flat \zeta_+
\bar F_\Lambda(\eta_-)  \right).
\ee
Eq.~\eqref{etapX} then
enables us to rewrite \eqref{eqwb0} in parallel to \eqref{eqva},
\be
\label{eqwb}
\txi_\Lambda = \tzeta_\Lambda
+ 2\I\, \e^{U + \frac12\cK(X,\bar X)}
\left( z \bar F_{\Lambda} + z^{-1} F_{\Lambda} \right).
\ee
Finally, using \eqref{etaA}, \eqref{etaB2-2} and \eqref{etasig},
one may show that 
\be
\label{eqaa0}
\alpha = \sigma + \zeta^\Lambda \txi_\Lambda - \tzeta_\Lambda \xi^\Lambda\,.
\ee
Together with \eqref{eqva} and \eqref{eqwb}, this implies
\be
\label{eqaa}
\alpha =
\sigma + 2\I\, \e^{U + \frac12\cK(X,\bar X)} \left(\bar W z + W z^{-1} \right)\,,
\ee
where $W$ is the symplectic invariant combination (or ``superpotential'')
\be
W =  F_\Lambda(X)\ \zeta^\Lambda - X^\Lambda \tzeta_\Lambda\,.
\ee

Altogether, \eqref{eqva}, \eqref{eqwb}, \eqref{eqaa} provide the general
relation between the complex coordinates $(\xi^\Lambda, \txi_\Lambda,
\alpha)$ on $\cZ$ and the real coordinates on the base $\cM$,
together with the fiber coordinate $z$.
Since it is one of the main results of this section, 
we rewrite the twistor map below:
\be
\label{gentwi}
\framebox{$
\begin{array}{ccc}
\xi^\Lambda &=& \zeta^\Lambda + 2\I\, \e^{U + \frac12\cK(X,\bar X)}
\left( z \bar X^{\Lambda} + z^{-1} X^{\Lambda} \right)\\
\txi_\Lambda &=& \tzeta_\Lambda + 2\I\, \e^{U + \frac12\cK(X,\bar X)}
\left( z \bar F_\Lambda +  z^{-1} F_\Lambda \right)\\
\alpha &=& \sigma + 2\I\, \e^{U + \frac12\cK(X,\bar X)}
\left(z \bar W  + z^{-1} W  \right)
\end{array}
$}
\ee

Finally we give a similar coordinate system in the \hk cone $\cS$:  
as discussed
in Section \ref{review}, we want two complex functions
$\pi^1$, $\pi^2$ on $\cS$, holomorphic in each fiber of $\cS$ over $\cM$,
defined up to the $\Z_2$ action $(\pi^1, \pi^2) \to (- \pi^1, -\pi^2)$,
and obeying $\pi^1 / \pi^2 = z$, $\chi = \abs{\pi^1}^2 + \abs{\pi^2}^2$.
A pair of coordinates satisfying our requirements is
\begin{equation} \label{hkc-coords}
\framebox{$\begin{pmatrix} \pi^1 \\ \pi^2 \end{pmatrix} = 2\,\e^U \sqrt{v^\flat}
\begin{pmatrix} z^\half \\ z^{-\half} \end{pmatrix}$}
\end{equation}
Indeed, we compute
\begin{equation}
\abs{\pi^1}^2 + \abs{\pi^2}^2 = 4\,\e^{2 U}\,\abs{v^\flat}\
(\abs{z} + \abs{z}^{-1})\,.
\end{equation}
Using $\abs{z} = \abs{\zeta_+}$ we see that this is $4 r^\flat \e^{2U}$,
which by \eqref{phichi} is equal to $\chi$ as desired. With the
knowledge of \eqref{hkc-coords}, we may then translate the holomorphic
contact form \eqref{contact-form} into the holomorphic trivialization
$v^\flat=1$ appropriate for comparison with \eqref{contact-form-local}, 
\be
X = 4\,e^{2U}\, \frac{\de z+ {\cal P}}{z} = \frac{\e^{K_\cZ}}
{1+z\bar z} \, \sqrt{\frac{\bar z}{z}}\, (\de z+{\cal P})\,.
\ee

\section{Integrability of the BPS geodesic flow \label{bpsgeo}}

In this section, we apply twistorial methods
to find the general solution for supersymmetric geodesic motion on
the quaternionic-K\"ahler metric \eqref{QK-metric}. After suitable
analytic continuation, this problem is equivalent
to the construction of stationary, spherically
symmetric black hole solutions in $\cN=2$ supergravity coupled to vector
multiplets \cite{Gunaydin:2005mx}, or
spherically symmetric instantons in $\cN=2$
supergravity coupled to hypermultiplets \cite{Gutperle:2000ve}.
The corresponding solutions (as well as their multi-centered generalizations)
have been known explicitly for some time 
\cite{Sabra:1997kq,Sabra:1997dh,Denef:2000nb,LopesCardoso:2000qm,deVroome:2006xu}.  
We rederive them here to illustrate the power of the twistor
formalism, and illuminate the geometric structure behind these supergravity 
solutions.
We expect that similar arguments can be used to generate new solutions in
a variety of other contexts where supersymmetry can be
reduced to holomorphy.

\subsection{Strategy}

As will be shown in \cite{gnpw-to-appear}, 
there is a correspondence between
geodesics on a
quaternionic-K\"ahler manifold $\cM$ and geodesics 
on its hyperk\"ahler cone ${\cS}$ with zero
angular momentum under the global $SU(2)$. Moreover,
BPS geodesic motion on $\cM$,
characterized by the condition that the quaternionic vielbein
$V^{AA'}$ pulled back to the geodesic has a right-eigenvector
with eigenvalue zero, 
is equivalent to holomorphic geodesic motion on ${\cS}$:
\be
\label{bpscond}
p_\aleph = 0\,,
\ee
where $p_{\aleph}$ denotes the canonical momenta
conjugate to the holomorphic coordinates $z^{\aleph}$ on $\cS$;
it will be convenient to choose the holomorphic coordinates
$(z^\aleph) = (\xi^\Lambda, \txi_\Lambda, w_\flat, v^\flat)$
on the hyperk\"ahler cone.  In particular, \eqref{bpscond} implies that the geodesic
is null,
\be \label{bpsnull}
p_{\aleph} g^{\aleph\bar\aleph} p_{\bar\aleph} = 0\,.
\ee
It is impossible for real non-constant geodesics on $\cS$ to satisfy \eqref{bpscond},
since $p_{\bar \aleph} = p_\aleph^*$.  For the analytic continuations of $\cS$ relevant
to the black hole or instanton problems, however, the BPS conditions can be satisfied.
In this section, we take the metric on $\cS$ to be the standard metric on the Swann
space of the quaternionic-K\"ahler $c$-map metric \eqref{QK-metric},
but treat the holomorphic and
anti-holomorphic coordinates $z^\aleph$ and $\bar{z}^{\bar\aleph}$ independently,
\textit{i.e.} we work with the complexification of $\cS$.  We return
to the issue of reality conditions at the end of Section \ref{sec-solution}.

The BPS condition \eqref{bpscond} implies that the anti-holomorphic
coordinates $(z^{\bar\aleph}) = (\bar\xi^\Lambda, \bar\txi_\Lambda,
\bar w_\flat, \bar v^\flat)$ are constants of motion.
Moreover, the conservation
of the Noether charges\footnote{$(P^\Lambda, Q_\Lambda, K)$ will in general
differ from the charges $(p^\Lambda,
q_\Lambda, k)$ on the base, \textit{e.g.} due
to the rescaling of the metric on ${\cal M}$
by $r^2=\chi$.} associated with the Heisenberg and $U(1) \subset SU(2)$ symmetries
(the latter vanishing for geodesics with zero $SU(2)$ momentum),
\bse
\bea
\label{conshol}
P^\Lambda &=& p_{\txi_\Lambda}+p_{\bar \txi_\Lambda} \,, \\
Q_\Lambda &=& - p_{\xi^\Lambda} -p_{\bar \xi^\Lambda} 
+\frac{\I}{2} ( \txi_\Lambda p_{w_\flat } 
- \bar \txi_\Lambda p_{\bar w_\flat } )\,, \\
K &=& \frac{\I}{4} ( p_{w_\flat }-p_{\bar w_\flat } )\,,  \\
0 &=& \frac{1}{2\I} ( v^\flat p_{v^\flat} - \bar v^\flat p_{\bar v^\flat} )\,,
\label{u1van}
\eea
\ese
implies that the anti-holomorphic momenta
$p_{\bar \aleph} = (p_{\bar\xi^\Lambda}, p_{\bar\txi_\Lambda},
p_{\bar w_\flat}, p_{\bar v^\flat})$
are also constants of motion (and, moreover, that
$p_{\bar v^{\flat}}=0$). It turns out that these conserved quantities
are sufficient to integrate the motion completely.

Indeed, $p_{\bar \aleph}$ being constant, the first order equation
\be
g_{\bar \aleph \aleph} \frac{\de z^\aleph}{\de t} = p_{\bar \aleph}
\ee
can be integrated using the K\"ahler property
$g_{\bar \aleph \aleph}= \pa_{z^{\bar \aleph}}\pa_{z^\aleph}\chi$
of the metric, to give
\be
\label{int1st}
\framebox{$
\pa_{\bar \aleph} \chi = p_{\bar \aleph} t + c_{\bar \aleph}
$}
\ee
where $c_{\bar \aleph}$ are constants of integration. Therefore,
in terms of the variables $\pa_{\bar \aleph} \chi$, the motion becomes linear.
This identifies the angle variables of this integrable system as the variables
conjugate to ${\bar z}^{\bar\aleph}$ under Legendre transform with respect to the
hyperk\"ahler potential $\chi$.  To find
the most general solution of BPS geodesic motion
on $\cS$, it only remains to express the complex variables $z^{\aleph}$ in
terms of  $\pa_{\bar \aleph} \chi$ and the constants of motion
$\bar{z}^{\bar \aleph},  p_{\bar \aleph}, c_{\bar \aleph}$.
Finally, the BPS geodesic motion can  be projected on the
quaternionic-K\"ahler base using the covariant $c$-map formulae of Section
\ref{sec-covcmap}, and enforcing the vanishing of
the $SU(2)$ momenta.

\subsection{Solution} \label{sec-solution}

We now exploit the explicit form \eqref{chihkc} of the hyperk\"ahler potential
for the $c$-map:
\be
\chi(\phi^\Lambda, \chi_{\Lambda}, \tilde \chi_\flat  )
= 2 \sqrt{v^\flat \bar v^\flat}\
\sqrt{\Sigma^2(\phi,\chi)-\tilde\chi_\flat ^2}\,,
\ee
where we recall that
\bea
\phi^\Lambda &=& \frac{\I}{2} (\xi^\Lambda-\bar \xi^\Lambda)\label{phiv}\,,
\qquad \chi_\Lambda = \frac{\I}{2} (\txi_\Lambda-\bar \txi_\Lambda)\,,\\
\tilde\chi_\flat  &=&
w_\flat +\bar w_\flat +\frac{\I}{4} (\xi^\Lambda+\bar \xi^\Lambda)
\label{wflatchi}
(\txi_\Lambda-\bar \txi_\Lambda)\,.
\eea
Using the identities
\be
\label{stabx}
\begin{pmatrix}
\pa_{\chi_\Lambda} \chi\\
 -\pa_{\phi^\Lambda} \chi
\end{pmatrix}
=\Re
\begin{pmatrix} \eta^\Lambda \\ F_\Lambda(\eta) \end{pmatrix}
=
\begin{pmatrix}  x^\Lambda - \frac{x^\flat}{2}
\left( \xi^\Lambda +  \bar \xi^\Lambda\right) \\
y_\Lambda - \frac{x^\flat}{2}
\left( \txi_\Lambda+\bar \txi_\Lambda\right)
 \end{pmatrix}
\ee
(where the second equality defines $y_\Lambda$) and
\be
\pa_{\tilde\chi_\flat}\chi = x^\flat \,,\quad
2 \I v^\flat \pa_{v^\flat} \chi =
\I \chi - (y_\Lambda-x^\flat \txi_\Lambda)\ \xi^\Lambda
\equiv  y_\flat\,,
\ee
where the partial derivatives of $\chi$ are taken in the
coordinates $(\phi^\Lambda,\chi_\Lambda,\tilde\chi_\flat)$, one may
rewrite the anti-holomorphic derivatives $\pa_{\bar z^{\bar\aleph}} \chi$
appearing in \eqref{int1st} as
\bea
\pa_{\bar\xi^\Lambda} \chi = \frac{\I}{2}
( y_\Lambda - x^\flat \bar\txi_\Lambda ) \ &,&\qquad
-\pa_{\bar\txi_\Lambda}\chi = \frac{\I}{2} x^\Lambda\,,  \\
\pa_{\bar w_\flat} \chi= x^\flat\ &,&\qquad \pa_{\bar v^\flat} \chi =
\frac{\chi}{2\bar v^\flat}\,.
\eea
Together with \eqref{int1st}, these identities imply that the
\hk potential $\chi$ is a constant of motion, while $x^\Lambda, y_\Lambda, x^\flat$ flow linearly:
\be
\label{xlin}
x^\Lambda =  2\I\,(P^\Lambda\,t + C^\Lambda)  \,,\quad
y_\Lambda =  2\I\, (Q_\Lambda\,t + D_\Lambda) \,,\quad
x^\flat = 4\I\,(K\,t+ E)\,.
\ee
It will be useful to further define
\be
\hat x^\Lambda \equiv x^\Lambda - x^\flat \bar \xi^\Lambda \,,\quad
\hat y_\Lambda \equiv y_\Lambda - x^\flat \bar \txi_\Lambda \,,
\ee
which, like $x^\Lambda$ and $y_\Lambda$, depend linearly on the geodesic time,
\be
\label{shiflin}
\hat x^\Lambda =  \chi(p^\Lambda\,t+c^\Lambda)  \,,\quad
\hat y_\Lambda =  \chi(q_\Lambda\,t + d_\Lambda) \,,\quad
x^\flat = \chi (k\,t+ e)\,,
\ee
with shifted and rescaled momenta,
\begin{gather}
\chi p^\Lambda=2\I P^\Lambda - 4\I K \bar \xi^\Lambda\,,\quad
\chi q_\Lambda=2\I Q_\Lambda - 4\I K \bar \txi_\Lambda\,,\quad
\chi k = 4\I K\,,\\
\chi c^\Lambda = 2\I C^\Lambda -4\I E \bar \xi^\Lambda\,,\quad
\chi d_\Lambda = 2\I D_\Lambda -4\I E \bar \txi_\Lambda\,,\quad
\chi e = 4\I E\,.
\end{gather}
In order to find the explicit trajectory, we note that $(\hat x^\Lambda,
\hat y_\Lambda)$ satisfy ``generalized stabilization equations''
analogous to \eqref{attr-charges},
\be
\label{genstab}
\frac12 \left[ C \eta^\Lambda_+ + \tilde C \eta^\Lambda_-  \right]
= r^\flat\,\hat x^\Lambda \,,\quad
\frac12 \left[ C F_\Lambda(\eta_+) + \tilde C \bar F_\Lambda(\eta_-) \right]
= r^\flat\,\hat y_\Lambda\,,
\ee
where $C$ and $\tilde C$ were defined in \eqref{ctc}.
Despite the fact that
$C$ and $\tilde C$ are in general not complex conjugate to one another,
the standard solution \eqref{solstab} to the stabilization equations continues
to hold,
\begin{subequations}
\label{fullsol}
\bea
C\eta^\Lambda &=& r^\flat \left( \hat x^\Lambda
+ \I \frac{\pa\Sigma(\hat x,\hat y)}{\pa \hat y_\Lambda}\right),  \qquad
\tilde C \bar \eta^\Lambda = r^\flat \left( \hat x^\Lambda
- \I \frac{\pa\Sigma(\hat x,\hat y)}{\pa \hat y_\Lambda} \right),  \\
C F_\Lambda &=& r^\flat \left( \hat y_\Lambda
- \I \frac{\pa\Sigma(\hat x,\hat y)}{\pa \hat x^\Lambda}\right),  \qquad
\tilde C \bar F_\Lambda  = r^\flat \left( \hat y_\Lambda
+ \I \frac{\pa\Sigma(\hat x,\hat y)}{\pa \hat x^\Lambda}\right).
\eea
\end{subequations}

Injecting these relations into \eqref{stabphi},
we may express $\phi^\Lambda$, $\chi_\Lambda$ in terms of
the linear flows $\hat x^\Lambda, \hat y_\Lambda$ as
\be \label{pclinear}
\begin{pmatrix}
\phi^\Lambda\\ \chi_\Lambda
\end{pmatrix}
=\frac{\I x^\flat}{C \tilde C}
\begin{pmatrix}
\hat x^\Lambda\\ \hat y_\Lambda
\end{pmatrix}
+ \frac{r^\flat}{C \tilde C}
\begin{pmatrix}
\frac{\pa\Sigma(\hat x,\hat y)}{\pa \hat y_\Lambda}\\
-\frac{\pa\Sigma(\hat x,\hat y)}{\pa \hat x^\Lambda}
\end{pmatrix}.
\ee
Since $(\phi^\Lambda,\chi_\Lambda)$ are related to the differences
$\xi^\Lambda-\bar\xi^\Lambda$ and $\txi_\Lambda-\bar\txi_\Lambda$ by \eqref{phiv}, and
since the anti-holomorphic coordinates $(\bar\xi^\Lambda, \bar\txi_\Lambda)$
are constants of motion, \eqref{pclinear} determines $(\xi^\Lambda, \txi_\Lambda)$ once $x^\flat$ and $r^\flat$ are known. The former is given by the linear flow
\eqref{shiflin}, while the latter follows from the
general property \eqref{entropy-charges} of the attractor equations,
\be
4 (r^\flat)^2\,\Sigma(\hat x,\hat y) = C\tilde C\,K(\eta_+,\bar\eta_+) =
16 \,(r^\flat) ^2\,v^\flat\bar v^\flat\,\Sigma(\phi,\chi)\,,
\ee
which, in combination with \eqref{chi-second}, leads to
\be \label{rflatsigma}
r^\flat = \frac{\Sigma(\hat x,\hat y)}{\chi}\,.
\ee
Finally, we may obtain the flows of $v^\flat$ and $\tilde\chi_\flat$ from (using \eqref{solx0})
\be
v^\flat = \frac{(r^\flat)^2 - (x^\flat)^2}{4 \bar{v}^\flat} \,,\quad
\tilde\chi_\flat = \frac{\chi \,x^\flat}{(r^\flat)^2 - (x^\flat)^2}\,,
\ee
and then infer the flow of $w_\flat$ from \eqref{wflatchi}.  We have now obtained all of the 
holomorphic coordinates $z^\aleph$, and hence determined the BPS geodesic trajectory 
on the \hk cone $\cS$.

In order to project the geodesic flow to the quaternionic-K\"ahler
base we use the covariant $c$-map formulae of Section \ref{sec-covcmap}.
In view of \eqref{etaz}, the evolution of the scalars $(z^a,\bar z^{\bar a})$ is
simply given by the ratios
\be
z^\Lambda = \frac{ \hat x^\Lambda
+ \I \frac{\pa\Sigma(\hat x,\hat y)}{\pa \hat y_\Lambda} }
{ \hat x^0
+ \I \frac{\pa\Sigma(\hat x,\hat y)}{\pa \hat y_0}}\,.
\ee
The evolution of the dilatonic variable $U$ follows from \eqref{phichi} and \eqref{rflatsigma},
\be
\label{fullu}
\e^{-2U} = \frac{4\,\Sigma(\hat x,\hat y)}{\chi^2}\,.
\ee
In the black hole context, the time $t$ is the inverse radial distance
in the spatial slices of the black hole, while the area of the sphere 
as a function of the radial distance is given by $4\pi \e^{-2U}/t^2$.
Using \eqref{shiflin}, the Bekenstein-Hawking entropy of the black hole is therefore given by
\be
\label{bhesse}
\framebox{$
S_{\rm BH}(p,q) = \lim_{t\to\infty} \pi \e^{-2U}/t^2  = 
4 \pi\,\Sigma(p^\Lambda,q_\Lambda)
$}
\ee
reproducing the known relation between black hole entropy and
Hesse potential \cite{Ooguri:2004zv,LopesCardoso:2006bg}.

The motion of $\zeta^\Lambda, \tzeta_\Lambda$ may be obtained
by rewriting \eqref{veta}, \eqref{eqwb0} in terms of $C,\tilde C$
and substituting \eqref{fullsol}, 
\bse
\label{xibarflow}
\bea
\zeta^\Lambda &=& \bar\xi^\Lambda - \frac{1}{2(r^\flat)^2}
\left[ C \eta_+^\Lambda - \tC \bar\eta_+^\Lambda \right]
=  \bar\xi^\Lambda - \frac{\I}{r^\flat}
\frac{\pa\Sigma(\hat x,\hat y)}{\pa \hat y_\Lambda}\,, \\
\tzeta_\Lambda &=& \bar\txi_\Lambda - \frac{1}{2(r^\flat)^2}
\left[ C F_\Lambda - \tC \bar F_\Lambda \right]
= \bar\txi_\Lambda + \frac{\I}{r^\flat}
\frac{\pa\Sigma(\hat x,\hat y)}{\pa \hat x^\Lambda}\,.
\eea
\ese
Finally, the flow of $\sigma$ follows from the complex conjugate of 
\eqref{eqaa0},
\be
\sigma =  \bar\alpha +\frac{\I}{r^\flat} \left[
{\bar \xi}^\Lambda \frac{\pa\Sigma(\hat x,\hat y)}{\pa \hat x^\Lambda}
+{\bar \txi}^\Lambda \frac{\pa\Sigma(\hat x,\hat y)}{\pa \hat y_\Lambda}
\right]\,.
\ee
It is also useful to note the solution for the holomorphic coordinates
$\xi^\Lambda,\txi_\Lambda$, obtained by similar manipulations:
\bse
\label{xiflow}
\bea
\xi^\Lambda &=& \zeta^\Lambda+\frac{1}{2 r^\flat}
\left[ \frac{C}{\tilde C} \left(
\hat x^\Lambda- \I \frac{\pa\Sigma(\hat x,\hat y)}{\pa \hat y_\Lambda}\right)
-\frac{\tilde C}{C} \left( 
\hat x^\Lambda+\I \frac{\pa\Sigma(\hat x,\hat y)}{\pa \hat y_\Lambda}\right)
\right],\\
\txi_\Lambda &=& \tzeta_\Lambda+\frac{1}{2 r^\flat}
\left[ \frac{C}{\tilde C} \left(
\hat y_\Lambda+ \I \frac{\pa\Sigma(\hat x,\hat y)}{\pa \hat x^\Lambda}\right)
-\frac{\tilde C}{C} \left( 
\hat y_\Lambda- \I \frac{\pa\Sigma(\hat x,\hat y)}{\pa \hat x^\Lambda}\right)
\right].
\eea
\ese

We recall that geodesic motion on ${\cS}$ only projects down to
geodesic motion on the base $\cM$ when the $SU(2)$ momentum 
vanishes.\footnote{More general geodesic motion on ${\cS}$ would descend
to motion on $\cM$ in a non-trivial magnetic field.}
The $U(1)\subset SU(2)$ charge has already been set to zero in
\eqref{u1van}, and $J_+$ vanishes for all holomorphic geodesics, but it 
remains to enforce
\be
J_- =
\hat x^\Lambda p_{\bar \xi^\Lambda}
- \hat y_\Lambda p_{\bar \txi_\Lambda}
- \frac{\I}{4}\bar y_\flat  p_{\bar w_\flat }
+ x^\flat p_{\bar v^\flat} = 0\,.
\ee
This determines the NUT charge $k$ as
\be
\label{nutcon}
\I k = p^\Lambda d_\Lambda - q_\Lambda c^\Lambda\,.
\ee
Thus, we have obtained the general BPS trajectory on the
complexification of the \qk metric \eqref{QK-metric}.

It remains to enforce the reality conditions appropriate to the problem at
hand. For BPS instantons, there
is no such reality condition, although one should in principle ensure
that the Euclidean configuration can be reached by analytic continuation
of the path integral.
For BPS black holes, $(U, \sigma)$ need to be real whereas $(\zeta^\Lambda,
\tzeta_\Lambda)$ need to be imaginary. This requires the charges $(P^\Lambda,
Q_\Lambda)$ to be imaginary  and $K$ real, while the situation is reversed
for $(p^\Lambda, q_\Lambda, k)$. Thus, $(x^\Lambda, y_\Lambda)$
need to be real while $x^\flat$ is imaginary and $r^\flat$ is real. 
Moreover, one should
demand that $(\bar\xi^\Lambda, \bar\txi_\Lambda)$ are imaginary and
$\bar\alpha$ is real.  
One may check from \eqref{xiflow}, \eqref{eqaa0} that this requires
that $(\xi^\Lambda, \txi_\Lambda)$
are imaginary while $\alpha$ is real.

We conclude that the 
twistor space for the para-\qk space $\cM^*$
is obtained by taking $(\xi^\Lambda, \txi_\Lambda, \bar\xi^\Lambda, 
\bar\txi_\Lambda)$ as independent, purely imaginary variables, and 
$(\alpha,\bar\alpha)$ as independent, real variables. We note that
then $z\bar z$ becomes a phase, as a consequence
of \eqref{zzbar}, and the
$S^2$ fiber of the twistor space becomes a hyperbolic two-plane $H_2$.

We may now compare our solution to the ones appearing in
\cite{Sabra:1997kq,Sabra:1997dh,Denef:2000nb,LopesCardoso:2000qm,Bates:2003vx,deVroome:2006xu}.
Using \eqref{defz}, we may express $(\eta^\Lambda_+,F_{\Lambda}(\eta_+))$
in terms of the symplectic section $(X^\Lambda, F_\Lambda(X))$ on
the special K\"ahler manifold $\cM_s$, and obtain
\be
-\I\, \bar z\,   \e^{U+\frac12 {\cal K}(X,\bar X)}
\begin{pmatrix}X^\Lambda \\ F_\Lambda \end{pmatrix}
+\I \, \bar z^{-1}\,   \e^{U+\frac12 {\cal K}(X,\bar X)}
\begin{pmatrix}\bar X^\Lambda \\ \bar F_\Lambda \end{pmatrix}
= \frac{1}{r^\flat}
\begin{pmatrix}\hat x^\Lambda \\ \hat y_\Lambda \end{pmatrix}.
\ee
Further expressing $r^\flat$ in terms of the dilaton using \eqref{phichi},
and setting $\bar z = 1/ (\bar z)^* =  \e^{-\I\beta}$, we find
 \be
\label{deneflin}
\Im\left[
 \e^{\half {\cal K}(X,\bar X)-U-\I\beta}
\begin{pmatrix} X^\Lambda \\ F_\Lambda \end{pmatrix}
\right]
=
\begin{pmatrix} H^\Lambda \\ H_\Lambda\end{pmatrix}
\ee
with
\be
\label{harmsph}
H^\Lambda = \hat x^\Lambda/ \chi = p^\Lambda t + c^\Lambda\,,\quad
H_\Lambda = \hat y_\Lambda / \chi = q_\Lambda t + d_\Lambda\,.
\ee
In the black hole problem, $t$ is identified with the inverse radial
distance on the $\IR^3$ spatial slices, so $(H^\Lambda,H_\Lambda)$ are
indeed harmonic functions, although not of the most general type
allowed in \cite{Sabra:1997kq,Sabra:1997dh,Denef:2000nb,LopesCardoso:2000qm}.
In particular,
we see that the phase $\beta$ (equal to the phase of the central charge
$Z$ near the horizon) is identified throughout the flow
as the azimuthal angle on the $S^2$ fiber of $\cZ$.
It would be very interesting to generalize
our discussion to the multi-centered case, and lift the standard
solutions to suitably holomorphic maps from $\IR^3$ to $\cS$.

\section{The quaternionic Penrose transform \label{pentran}}

The classic Penrose transform relates wave functions on subsets of 
the twistor space
$\C\PP^3$ --- more precisely, elements in the sheaf cohomology 
$H^1(X \subset \C\PP^3, \cO(-2h-2))$ --- to
helicity-$h$ solutions of conformally invariant wave equations on 
subsets of $\R^4$ (see \textit{e.g.} \cite{MR944085,ward-wells,MR1292461}).
This transform has been generalized in many directions.  For example, 
one may replace 
$\R^4$ by another self-dual $4$-manifold \cite{MR563832,MR634238}.  
Self-dual $4$-manifolds can be
considered as the $n=1$ case of \qk $4n$-manifolds, and there is a 
further extension of the
Penrose transform to this case \cite{quatman,MR1165872}.
Letting $\cM$ be a \qk manifold and $\cZ$ its twistor space, this 
``quaternionic Penrose transform''
relates elements in $H^1(Y \subset \cZ, \cO(-k))$ to solutions of wave
equations constructed from the quaternionic structure on subsets of $\cM$.

Here we work out one aspect of this transform explicitly:  we represent elements
$\psi_f \in H^1(Y \subset \cZ, \cO(-k))$ by holomorphic sections $f$ on an 
appropriate open set in $Y$, and we
give a contour integral formula which transforms any such $f$ into a solution of
a wave equation on $\cM$.
Furthermore, in the case where $\cM$ comes from the $c$-map,
we use the results of Section \ref{sec-twistor-map}
to make this transform particularly concrete:  it converts holomorphic
functions in $2n+1$ variables, $g(\xi^\Lambda, \txi_\Lambda, \alpha)$,
to solutions of wave equations on $\cM$.  Finally we use this formalism 
to compute the Penrose transform of particular eigenstates of the
Heisenberg group; physically, these are interpreted
as the wave function of BPS black holes in radial 
quantization \cite{gnpw-to-appear}.

\subsection{For general quaternionic-\kahler manifolds} \label{penrose-general}

Suppose $\cM$ is a quaternionic-\kahler manifold.  As we reviewed in Section \ref{twiswa},
there is a natural $\C\PP^1$-bundle over $\cM$, the twistor space $\cZ$,
equipped with a canonical complex structure.
The Penrose transform is local on $\cM$, so we may as well take $\cM$ to be a small open set;
in particular, we may assume that the decomposition $T_\C \cM = E \otimes H$ exists globally
on $\cM$.  Then $\cZ$ is equipped with a canonical holomorphic line bundle $\cO(1)$.
We choose a standard local trivialization of $H$, and let $(\pi^1, \pi^2)$ be the
corresponding coordinates on the total space of $H^\times$.
Then $z = \pi^1 / \pi^2$ is a coordinate on the $\C\PP^1$ fibers of $\cZ$.

To construct the Penrose transform
we begin by constructing a class $\psi_f$ in the sheaf cohomology $H^1(\cZ, \cO(-k))$.  
The cohomological interpretation of the Penrose transform has been described 
in \cite{Eastwood:1981jy,ward-wells};
we construct $\psi_f$ using the \cech description of the cohomology, 
reviewed \textit{e.g.} in \cite{MR95d:14001,ward-wells}.
The \cech construction depends on a covering of $\cZ$ by open sets:  
we take two sets, $U_1 = \{(x,z): x \in \cM, z \neq 0\}$ and $U_2 = \{(x,z): x \in \cM, z \neq \infty\}$.
Then $\psi_f$ is represented simply by a holomorphic section $f$ of
$\cO(-k)$ on $U_1 \cap U_2$.  Equivalently, we may regard $f$ as a
function $f(x, \pi^1, \pi^2)$, homogeneous of degree $-k$ in the $\pi^{A'}$,
defined where $\pi^1 \neq 0$, $\pi^2 \neq 0$, and holomorphic.  Here 
``holomorphic''
is defined using the complex structure on the total space of $\cO(-1) \to \cZ$, 
described in Section \ref{review}:  concretely, it implies that all the 
vector fields
\begin{equation} \label{defd}
d_A \equiv \pi^{A'} \partial_{A A'} - \pi^{B'} \pi^{C'} (p_{A B'})^{D'}_{C'} \frac{\pa}{\pa \pi^{D'}}
\end{equation}
annihilate $f$.  (To check this, one shows that each $d_A$ has zero 
inner product with the basis of $(1,0)$-forms on $\cS$ described in Section \ref{review},
so it is a $(0,1)$ vector field, \textit{i.e.} an antiholomorphic derivative.)

We now construct the Penrose transform $\varphi$ of $\psi_f$ as an 
appropriate contour 
integral of $f$.  For $k>2$, we will show by simple manipulations that 
the holomorphy of $f$ implies 
$\varphi$ obeys first-order differential 
equations on $\cM$.  We then turn to the $k=2$ case,
which leads to second-order differential equations and is technically 
more difficult.
The Penrose transform for $k<2$ will involve in general differentiation as
well as integration, but should be treatable along the same lines as in
the classic case; we do not consider it here.

\subsubsection*{For $\cO(-k)$, $k > 2$:}

For notational simplicity we treat mainly the case $k=3$.
So given $f$ representing $\psi_f \in H^1(\cZ, \cO(-3))$, we construct a field on $\cM$ by
\begin{equation} \label{twistor-ci-O3}
\framebox{$
\varphi^{A'}(x) = \oint \left( \pi_{B'} \de \pi^{B'} \right) \pi^{A'} f(x, \pi)
$}
\end{equation}
Since $f$ has homogeneity $-3$ and there are $3$ explicit factors of $\pi$,
the whole integrand has homogeneity $0$, so it is well defined on $\cZ$.
The contour of integration is chosen to lie in the $\C\PP^1$ fiber of $\cZ$ over $x$, and to separate
$z=0$ and $z=\infty$.

In the rest of this section, we will prove that $\varphi^{A'}(x)$ obeys a Dirac-type equation,
\begin{equation} \label{dirac}
\nabla_{AA'} \varphi^{A'}(x) = 0\,.
\end{equation}
The strategy is simple:  because $f$ is
holomorphic on $\cZ$ we have $d_A f = 0$.
We insert this into the contour integral \eqref{twistor-ci-O3} to get
\begin{equation} \label{twistor-annih}
0 = \oint \left( \pi_{B'} \de\pi^{B'} \right) \left(\pi^{A'} \p_{A A'} - \pi^{E'} \pi^{A'} (p_{A E'})^{G'}_{A'}   \frac{\pa}{\pa\pi^{G'}} \right) f(x, \pi)\,.
\end{equation}
Now integrate the operator $\frac{\pa}{\pa\pi^{G'}}$ by
parts, using the identity (easily checked in local coordinates)
\begin{equation}
\oint  \left( \pi_{B'} \de\pi^{B'} \right) \frac{\pa}{\pa \pi^{G'}} g(\pi) = 0\,.
\end{equation}
Applied to \eqref{twistor-annih} this
integration by parts gives two terms, since $\frac{\pa}{\pa \pi^{G'}}$ can hit either $\pi^{E'}$ or $\pi^{A'}$.
If it hits $A'$ we get $(p_{A E'})^{A'}_{A'}$ which vanishes; so we only get the $E'$ term, giving
\begin{align} \label{twistor-annih-2}
0 &= \oint \left( \pi_{B'} \de\pi^{B'} \right)  \pi^{A'} \left(\p_{A A'} + (p_{A E'})^{E'}_{A'} \right) f(x, \pi)\,.
\end{align}
The right side is $\nabla_{A A'} \varphi^{A'}$, so we get the desired \eqref{dirac}.

More generally, the twistor transform for $\cO(-k)$ with $k>2$ gives totally symmetric $(k-2)$-tensors on $\cM$;
it is obtained by replacing \eqref{twistor-ci-O3} with
\begin{equation} \label{twistor-ci-Ok}
\varphi^{(A'_1 A'_2 \cdots A'_{k-2})}(x) = \oint \left( \pi_{B'} \de\pi^{B'} \right) \pi^{A'_1} \pi^{A'_2} \cdots \pi^{A'_{k-2}} f(x, \pi)\,,
\end{equation}
and the same differentiation under the integral sign we did above shows
\begin{equation}
\framebox{$
\nabla_{A A'_1}\, \varphi^{(A'_1 A'_2 \cdots A'_{k-2})}(x) = 0
$}
\end{equation}

\subsubsection*{For $\cO(-2)$:}

Now we turn to the harder case $k = 2$.
Given $f$ representing $\psi_f \in H^1(\cZ, \cO(-2))$, we construct a scalar function on $\cM$ by
a contour integral similar to \eqref{twistor-ci-O3},
\begin{equation} \label{twistor-ci-O2}
\framebox{$
\varphi(x) = \oint \left( \pi_{B'} \de\pi^{B'} \right) f(x, \pi)
$}
\end{equation}

In the rest of this section we prove that $\varphi(x)$ obeys a family of second-order differential equations,
\begin{equation} \label{qk-laplacian}
\left( \nabla_{AA'} \nabla_B^{A'} - \nu \eps_{AB} \right) \varphi(x) = 0\,,
\end{equation}
where we recall that $\nu = \frac{1}{4n(n+2)} R$.
Quaternionic geometry in case $n=1$ reduces to conformal geometry, and correspondingly
\eqref{qk-laplacian} reduces to the conformal Laplacian $\Delta - \frac{1}{6}R$ in that case.
For $n>1$, \eqref{qk-laplacian} gives more than one equation, transforming in $\wedge^2(E)$;
tracing them with $\eps^{AB}$ gives $\Delta - \frac{1}{2(n+2)} R$,
which differs from the conformal Laplacian $\Delta - \frac{4n-2}{4(4n-1)} R$.

To establish \eqref{qk-laplacian} we begin by
defining a second differential operator $d'_B$ on $\cS$,
acting on sections $f_A(x,\pi)$ by
\begin{equation} \label{defdp}
d'_B f_A = \left( \pi^{A'} \partial_{B A'} - \pi^{B'} \pi^{C'} (p_{B B'})^{D'}_{C'} \frac{\pa}{\pa \pi^{D'}} \right) f_A + \pi^{A'} (q_{B A'})^D_A f_D\,.
\end{equation}
This operator is engineered to obey
\begin{equation} \label{twistor-complex}
d'_A d_B - d'_B d_A = 0\,.
\end{equation}
To prove this, we choose a local frame such that $p = 0$ at $x$:  then
\eqref{twistor-complex} is
\begin{equation} \label{twistor-complex-2}
\pi^{A'} \pi^{B'} \pi^{C'} \left( \pa_{A A'} (p_{B B'})_{C'}^{D'} - \pa_{B A'} (p_{A B'})_{C'}^{D'} \right) \frac{\pa}{\pa \pi^{D'}} = 0\,.
\end{equation}
On the other hand, in these coordinates the formula \eqref{usp2-curvature} for the $USp(2)$ curvature becomes
\begin{equation} \label{h-curvature-normal}
\pa_{AA'} (p_{BB'})^{D'}_{C'} - \pa_{BB'} (p_{AA'})^{D'}_{C'} = \frac{\nu}{2} \eps_{AB} (\delta^{D'}_{A'} \eps_{C'B'} + \delta^{D'}_{B'} \eps_{C'A'})\,.
\end{equation}
This vanishes when symmetrized over $(A'B'C')$, establishing \eqref{twistor-complex-2}.

As we will now see, the complex
\begin{equation}
\cO(-2) \stackrel{d_A}{\rightarrow} \cO_A(-1) \stackrel{d'_B}{\rightarrow} \cO_{AB}(0)
\end{equation}
on $\cZ$ leads to a differential equation on $\cM$, which will turn out to be \eqref{qk-laplacian}.
Abstractly this equation arises as one of the differentials in a spectral sequence computing
$H^*(\cZ, \cO(-2))$, as sketched in \cite{MR1165872}, along the lines
of similar arguments described in \cite{Eastwood:1981jy, MR1038279}; but in this case
the construction is simple enough that it can be worked out by hand.
We begin by considering the function
\begin{equation}
h(x,\pi) \equiv \frac{\varphi(x)}{\pi^1 \pi^2} - f(x,\pi)
\end{equation}
on $\cS$.  By construction, contour-integrating $h$ gives
\begin{equation}
\oint \left( \pi_{B'} \de\pi^{B'} \right) h(x, \pi) = 0\,.
\end{equation}
But since $H^1(\C\PP^1, \cO(-2))$ is one-dimensional,
this vanishing implies that $h(x,\pi)$ is trivial in $H^1(\C\PP^1, \cO(-2))$; in other words,
there is a decomposition
$h = h^{(1)} + h^{(2)}$
where $h^{(i)}$ is defined on $U_i$.
Applying $d_A$ to this,
\begin{equation}
d_A h = d_A h^{(1)} + d_A h^{(2)}\,.
\end{equation}
Moreover, this is the unique decomposition of $d_A h$ into a piece defined on $U_1$ and a piece defined
on $U_2$:  the uniqueness follows from the fact that the ambiguity would be a global section of $\cO(-1)$
over $\C\PP^1$, and there are no such sections.  We now compute this decomposition in another way:
using $d_A f = 0$ we see that
$d_A h = d_A \left( \frac{\varphi}{\pi^1 \pi^2} \right)$, and using the definition \eqref{defd} of $d_A$ this gives
\begin{equation} \label{decomp-dah}
d_A h = {\frac{1}{\pi^1} \left[ \partial_{A2} + (p_{A1})^1_2 + \frac{\pi^2}{\pi^1} (p_{A2})^1_2 \right]\varphi} + {\frac{1}{\pi^2} \left[ \partial_{A1} + (p_{A2})^2_1 + \frac{\pi^1}{\pi^2} (p_{A1})^2_1 \right]\varphi}\,.
\end{equation}
The first (resp. second) term is defined on $U_1$ (resp. $U_2$), so by the uniqueness of the decomposition,
\begin{equation} \label{dah}
d_A h^{(1)} = \frac{1}{\pi^1} \left[ \partial_{A2} + (p_{A1})^1_2 + \frac{\pi^2}{\pi^1} (p_{A2})^1_2 \right]\varphi\,.
\end{equation}
Then using \eqref{twistor-complex},
\begin{equation} \label{twistor-complex-applied}
(d'_A d_B - d'_B d_A) h^{(1)} = 0\,,
\end{equation}
gives a second-order differential equation for $\varphi$.  To work out this equation it is again
convenient to work in normal coordinates on $\cM$, with $p = q = 0$ at $x$.  Then substituting the definition \eqref{defdp} of $d'_B$ and \eqref{dah} in \eqref{twistor-complex-applied}, the terms proportional to $\pi^2 / \pi^1$ vanish using \eqref{h-curvature-normal}, leaving
\begin{equation} \label{noncovariant-do}
\left( \pa_{A1} \pa_{B2} - \pa_{B1} \pa_{A2} + \pa_{A1} (p_{B1})_2^1 - \pa_{B1} (p_{A1})_2^1 \right) \varphi(x) = 0\,.
\end{equation}
This is not written in a manifestly $USp(2)$-covariant way, which can be traced back to the fact that our
covering of $\cZ$ by $\{U_1, U_2\}$ is not covariant.
Using \eqref{h-curvature-normal}, it is easily rewritten in the desired form,
\begin{equation}
\framebox{$
\left( \nabla_{AA'} \nabla_B^{A'} - \nu\, \eps_{AB} \right) \varphi(x) = 0
$}
\end{equation}

\subsection{For $c$-map spaces}

This general construction can be made more explicit when the quaternionic-\kahler space $\cM$
arises from the $c$-map:  as we will see, in this case the Penrose transform allows us to
construct solutions of wave equations on $\cM$
starting from arbitrary holomorphic functions in $2n+1$ variables
$g(\xi^\Lambda, \txi_\Lambda, \alpha)$.

So suppose $\cM$ comes from the $c$-map.  Then we have local complex coordinates $(\xi, \txi, \alpha)$
which cover an open set in $\cZ$.  To be precise, using \eqref{gentwi}, 
we see that this coordinate system covers all of each twistor sphere except the north and south
poles.  We also have a natural trivialization of $\cO(-k)$,
provided by the $\C^\times$ gauge condition $v^\flat = 1$, which enables us to pass between
homogeneous functions on $\cS$ and ordinary functions of $(\xi^\Lambda, \txi_\Lambda, \alpha)$.
So in this gauge a class in $H^1(\cZ, \cO(-k))$ can be simply represented by 
a holomorphic function $g(\xi^\Lambda, \txi_\Lambda, \alpha)$, while
the integration measure in \eqref{twistor-ci-O2} is obtained 
by using \eqref{hkc-coords}
in the $v^\flat = 1$ gauge. Thus, the Penrose transform for
scalar fields \eqref{twistor-ci-O2}
becomes
\begin{equation} \label{twistor-ci-O2-cmap}
\varphi(U, z^a, \bar z^{\bar a}, \zeta^\Lambda, \tzeta_\Lambda, \sigma) =
4 \  \e^{2 U}\  \oint \frac{\de z}{z}\,g(\xi^\Lambda, \txi_\Lambda, \alpha)\,,
\end{equation}
where $\xi^\Lambda$, $\txi_\Lambda$, $\alpha$ are given in terms of
$(U, X, \zeta^\Lambda, \tzeta_\Lambda, \sigma)$ by \eqref{gentwi}.
Similarly \eqref{twistor-ci-Ok} becomes
\begin{equation} \label{twistor-ci-Ok-cmap}
\framebox{$\varphi^{(A'_1 A'_2 \cdots A'_{k-2})}(U,  z^a, \bar z^{\bar a},
\zeta^\Lambda, \tzeta_\Lambda, \sigma) = 2^k\  \e^{k U}\
\oint \frac{\de z}{z}\,z^{\half \delta} g(\xi^\Lambda, \txi_\Lambda, \alpha)
$}
\end{equation}
where $\delta \equiv$ ((the number of $i$ with $A'_i = 1$) $-$ (the number of $i$ with $A'_i = 2$)).

When $k$ is odd, \eqref{twistor-ci-Ok-cmap} involves square roots of $z$.
This is related to the fact that $\cS$ provides a global definition of $\cO(k)$ and $S^k(H)$ only for $k$
even.  It is not obvious whether one can make sense of \eqref{twistor-ci-Ok-cmap} for
$k$ odd (perhaps by choosing $g$ with some appropriate branch cuts).

\subsection{The inner product}

Since we view $\psi \in H^1(\cZ, \cO(-k))$ as a wave function, it is natural 
to ask whether there is a canonical inner product $\inprod{\psi \vert \psi'}$ 
defined in terms of the geometry of $\cZ$.  For the classical case where $\cZ$ is a subset 
of $\IC\IP^3$, the answer to this question can be phrased in terms of an isomorphism 
$H^1(\cZ, \cO(-k)) \simeq H^1(\cZ, \cO(k-4))$ known as the 
``twistor transform'' \cite{MR610183}. Upon representing the classes
in $H^1$ by holomorphic functions, the corresponding inner product 
admits a concrete integral representation, given \textit{e.g.} in Section 3.3 of
\cite{Penrose:1972ia}.
While we do not know the generalization of the twistor transform 
to the $c$-map case, we can construct a candidate for
an inner product,
\begin{equation}
\inprod{\psi_f \vert \psi_{f'}} = \int_{\cZ'} \mathrm{vol}_\cZ\,\inprod{f \vert f'}
\end{equation} 
Here, $\inprod{f \vert f'}$ is the Hermitian inner product in $\cO(-k)$ 
and $\mathrm{vol}_\cZ$ the volume form induced from the \kahler-Einstein 
metric on $\cZ$.  This formula is not well defined \textit{a priori}; it only makes 
sense under an assumption about the global structure of $\cZ$, namely, 
every $\psi$ is obtained as $\psi_f$, with $f$ a unique
holomorphic section of $\cO(-k)$ over $\cZ' = \{ z \neq 0, z \neq \infty\} \subset \cZ$. 
There is some evidence that this assumption does hold when $\cM$ 
is a symmetric space \cite{gnpw-to-appear-2}.

Choosing the $v^\flat = 1$ gauge
to trivialize $\cO(-k)$, $f$ and $f'$ become $g(\xi^\Lambda, \txi_\Lambda, \alpha)$ and $g'(\xi^\Lambda, \txi_\Lambda, \alpha)$ as above, so
\begin{equation}
\inprod{f \vert f'} = \bar{g}\, g'\, \e^{k K_\cZ}
\end{equation}
where $K_\cZ$ is given in \eqref{kz-hesse}.
We determine the volume form by considering the line bundle $K$ of holomorphic top-forms on $\cZ$.
$K$ admits a natural Hermitian metric, 
in which the squared norm of any $\omega$ is $\frac{\omega \wedge \bar{\omega}}{\mathrm{vol}_\cZ}$.  
On the other hand $K \simeq \cO(-2n-2)$, and the metric is the $(2n+2)$-th power of the 
metric in $\cO(-1)$ \cite{MR664330}.  Recall from Section \ref{review} that the squared norm of 
the Heisenberg invariant section $v^\flat = 1$ is by definition $\e^{K_\cZ}$; hence
the Heisenberg invariant section $\de \xi^\Lambda \de \txi_\Lambda \de \alpha$ 
has squared norm $\e^{(2n+2) K_\cZ}$, up to an overall constant.
Comparing these two gives
\begin{equation}
\mathrm{vol}_\cZ = \e^{-(2n+2) K_\cZ}\,\de \xi^\Lambda \de \txi_\Lambda \de \alpha \, \de \bar{\xi}^\Lambda \de \bar{\txi}_\Lambda \de \bar{\alpha}\,.
\end{equation}
So altogether we find the inner product of wave functions as
\begin{equation}
\label{inner}
\framebox{$
\inprod{\psi_f \vert \psi_{f'}} = \int_{\cZ'} 
\de \xi^\Lambda \de \txi_\Lambda \de \alpha \, \de \bar{\xi}^\Lambda \de \bar{\txi}_\Lambda \de \bar{\alpha}\ \e^{(k-2n-2) K_\cZ}\,\overline{g(\xi^\Lambda, \txi_\Lambda, \alpha)}\, g'(\xi^\Lambda, \txi_\Lambda, \alpha)
$}
\end{equation}
$\cZ'$ does not cover the full range of the complex coordinates $(\xi^\Lambda,\txi_\Lambda,\alpha)$:  
the integration should run over a domain such that the bracket in \eqref{kz-hesse} is strictly positive.

\subsection{The BPS black hole wave function}

As an example of this technology, we compute the Penrose transform
of a class in $H^1(\cZ, \cO(-k))$ $(k \ge 2)$ which is an eigenvector
for the Heisenberg group acting on $\cZ$, with vanishing central character.
As will be explained in \cite{gnpw-to-appear}, such a class
describes the wave function of a BPS black hole with
fixed real electric and magnetic charges $(q_\Lambda, p^\Lambda)$ and
vanishing NUT charge, in a mini-superspace radial quantization scheme.

Given the action \eqref{heisz} of the complexified Heisenberg algebra
on the twistor space $\cZ$, an eigenvector
is determined up to normalization by its eigenvalues 
$\I p^\Lambda, \I q_\Lambda$ under $P^\Lambda$ and $Q_\Lambda$,
\begin{equation}
\label{cohst}
g(\xi^\Lambda, \txi_\Lambda, \alpha) =  \e^{\I( p^\Lambda \txi_\Lambda-q_\Lambda \xi^\Lambda)}\,.
\end{equation}
We expect that this wave function 
is delta-function normalizable with respect to the
inner product \eqref{inner} (perhaps after regulating by appropriately continuing $k$.)  
In physical applications, one would expect to
consider a quotient of $\cZ$ by a lattice in the Heisenberg group, which would select
integer momenta $p^I,q_I$; the wave function \eqref{cohst} then
should become normalizable, as the flat directions $\xi^\Lambda
-\bar\xi^\Lambda$, $\txi^\Lambda
-\bar\txi^\Lambda$, $\alpha-\bar\alpha$ become compact.

We now compute the Penrose transform of \eqref{cohst}, starting with
the case $k=2$. Equation \eqref{twistor-ci-O2-cmap} gives
\begin{equation}
\varphi(U, z^a, \bar z^{\bar a},\zeta^\Lambda, \tzeta_\Lambda, \sigma) = 4  \e^{2U + \I p^\Lambda \tzeta_\Lambda - \I q_\Lambda \zeta^\Lambda} \oint \frac{\de z}{z} \exp \left( -  \e^{U} (z \bar{Z} + z^{-1} Z) \right),
\end{equation}
where
\begin{equation}
Z =  \e^{\half \cK(X,\bar X)} (p^\Lambda F_\Lambda(X) 
- q_\Lambda X^\Lambda)\,
\end{equation}
is the ``central charge'', familiar from $\cN=2$ supergravity.
Using $\oint \frac{\de z}{z}  \e^{az + bz^{-1}} = 2 \pi I_0(2\sqrt{ab})$ (the modified Bessel function),
\begin{equation}
\varphi(U, z^a, \bar z^{\bar a}, \zeta^\Lambda, \tzeta_\Lambda, \sigma) = 8 \pi\, \e^{2U + \I p^\Lambda \tzeta_\Lambda - \I q_\Lambda \zeta^\Lambda}\ I_0(2  \e^U \abs{Z})\,.
\end{equation}
More generally for $k \ge 2$, \eqref{twistor-ci-O2-cmap} or \eqref{twistor-ci-Ok-cmap} give
\begin{equation}
\varphi^{(A'_1 A'_2 \cdots A'_{k-2})}(U, z^a, \bar z^{\bar a}, \zeta^\Lambda, \tzeta_\Lambda, \sigma) = 2^k\, \e^{kU + \I p^\Lambda \tzeta_\Lambda - \I q_\Lambda \zeta^\Lambda} \oint \frac{\de z}{z} z^{\half \delta} \exp \left( -  \e^{U} (z \bar{Z} + z^{-1} Z) \right),
\end{equation}
and using $\oint \frac{\de z}{z} z^m \e^{az + bz^{-1}} = 2 \pi\,\left(\frac{a}{b}\right)^{\frac{m}{2}} I_{-m} (2\sqrt{ab})$\,,
\begin{equation}
\label{bpswf}
\framebox{$
\varphi^{(A'_1 A'_2 \cdots A'_{k-2})}(U, z^a, \bar z^{\bar a}, \zeta^\Lambda, \tzeta_\Lambda, \sigma) = 2^{k+1} \pi\, \e^{k U + \I p^\Lambda \tzeta_\Lambda - \I q_\Lambda \zeta^\Lambda} \left( \frac{\bar Z}{Z} \right)^{\frac{\delta}{4}} I_{-\frac{\delta}{2}} (2  \e^U \abs{Z})
$}
\end{equation}
at least when $k$ (and hence $\delta$) is even. Irrespective of the value 
of $k$, we see that in the ``weak coupling'' or ``near horizon'' 
limit $U\to -\infty$, the wave function $\varphi$
as a function on the special K\"ahler manifold 
has minima at the minima of $|Z|$, and grows exponentially away from these
points. In the application to black hole physics, however, the required 
analytic continuation of the charges turns the modified
Bessel function $I$ into a $J$ Bessel function, with phase
stationary at the stationary points of the central charge $|Z|$, 
and modulus power-suppressed away from these points. 
This is consistent with the classical attractor 
behavior \cite{Ferrara:1995ih,Ferrara:1996um,Ferrara:1997tw},
although the absence of exponential decay is perhaps unexpected.
We shall return to the
physical interpretation of this wave function in the black hole context
in \cite{gnpw-to-appear}, and in the instanton
context in \cite{npv-to-appear}.

\section*{Acknowledgements}

We are grateful to M.~Ro\v{c}ek and A.~Waldron
for collaboration at an early stage of this project,
and to M.~G\"unaydin for many discussions on quasi-conformal realizations.
We also thank C.~LeBrun for useful explanations.
This research is supported in part by  the EU under contracts
MTRN--CT--2004--005104, MTRN--CT--2004--512194, by
ANR (CNRS--USAR) contract No 05--BLAN--0079--01, and by INTAS contract
03-51-6346.  The research of
A.~N. is supported by the Martin A. and Helen Chooljian Membership at the
Institute for Advanced Study and by NSF grant PHY-0503584.

\appendix

\section{Details on the superconformal quotient}

In this Appendix, we give some more details and useful formulas to
prove the results discussed in Section \ref{sec-covcmap}.  The superconformal
quotient can be performed either on the tensor multiplet side
or on the hypermultiplet side: we choose the former. In
the notation of \cite{Robles-Llana:2006ez} (but with $\cK\rightarrow -\cK$) 
the relevant bosonic
terms of the tensor multiplet Lagrangian coupled to Poincar\'e
supergravity after the $c$-map read
\begin{equation}\label{cmap-lagr}
\begin{split}
e^{-1}{L} = & \, - \frac{1}{2}
(\partial_\mu\phi)^2 - 2 \, {\mathcal G}_{a \bar b}\partial_\mu
z^a \partial^\mu \zb^{\bar b} + \frac{1}{2}{\e}^{-\phi}
({\mathcal N}+\bar{\mathcal N})_{\La\Si}
\partial_\mu A^\La\partial^\mu A^\Si \\
& \, + 2 \, \cT_{IJ} \, E_\mu^I E^{J\mu} + \I({\mathcal
N}-\bar{\mathcal N})_{\La\Si}\left[(\partial_\mu
A^\La)E^{\Si\mu}-2(\partial_\mu A^\La)A^\Si E^{0\mu}\right].
\end{split}
\end{equation}
Here $E^{\mu} = \tfrac{\I}{2} e^{-1} \varepsilon^{\mu \nu \rho
\sigma}
\partial_\nu E_{\rho \sigma}$ is the field strength of the
antisymmetric tensor field $E_{\mu \nu}$. The index $I$ runs over
one more value than $\Lambda$, so $I=\{\flat,\Lambda\}$. The matrix
$\cT_{IJ}$ appearing in the tensor field kinetic term is given by
\begin{equation}\label{defT}
\cT_{IJ}  = {\e}^{\phi} \left[
\begin{array}{cc} {\e}^{\phi} - ({\mathcal N}+\bar {\mathcal
N})_{\La\Si}A^\La
A^\Si & \frac{1}{2}({\mathcal N}+\bar{\mathcal N})_{\La\Si}A^\La \\
\frac{1}{2}({\mathcal N}+ \bar{\mathcal N})_{\La\Si}A^\Si &
-\frac{1}{4}({\mathcal N}+\bar{\mathcal
N})_{\La\Si}\end{array}\right],
\end{equation}
where ${\cal N}$ is defined as in \eqref{def-curlyN}. The relation
between these variables and those of the main text is given in 
footnote \ref{rvv-vs-us} on page \pageref{rvv-vs-us}.

Our task is to prove that the Lagrangian \eqref{cmap-lagr} follows
from the superspace Lagrangian density ${\cal L}$ given in \eqref{cont-cmap}.
The component Lagrangian for the rigid superconformal supersymmetric tensor
fields follows from \cite{Hitchin:1986ea,deWit:2001dj}. The relevant terms of
the bosonic Lagrangian read
\begin{eqnarray}\label{conf-lagr}
L&=&{\cal L}_{x^Ix^J}\Big(-\frac{1}{4}\partial_\mu x^I \partial^\mu x^J
-\partial_\mu v^I\partial^\mu {\bar v}^J+E_\mu^I E^{\mu J}\Big)\nonumber\\
&&- \I E_\mu^I\Big({\cal L}_{v^Ix^J}\partial^\mu v^J - {\cal L}_{{\bar v}^I
x^J}\partial^\mu {\bar v}^J\Big).
\end{eqnarray}
The Lagrangian \eqref{conf-lagr} has conformal symmetry. The scaling weight
of the scalars is $2$ and the matrix of second derivatives of ${\cal L}$
has scaling weight minus two. The function ${\cal L}$ itself has therefore
scaling weight plus two. From now on, we denote ${\cal L}_{IJ}\equiv
{\cal L}_{x^Ix^J}$. To obtain the Poincar\'e theory for the tensor
multiplets, one first couples
to the Weyl multiplet and integrates out the $SU(2)$ gauge fields. This
procedure is called the superconformal quotient and
was carried out in \cite{deWit:2006gn,Robles-Llana:2006ez}.
We will here apply it to the case of the $c$-map. It suffices to
consider only the terms quadratic in the tensor fields, the rest is fixed
by supersymmetry. Following \cite{deWit:2006gn,Robles-Llana:2006ez},
the zero weight matrix that multiplies the two tensors in the
Poincar\'e theory is
\begin{equation}
e^{-1}L_{{\rm TT}}={\cal H}_{IJ}E_\mu^IE^{\mu J}\,,
\end{equation}
with
\begin{equation}\label{HIJ}
{\cal H}_{IJ}=\chi_{\rm T}\Big({\cal L}_{IJ}+\frac{1}{4}
{\cal L}_{IK}(\vec {r}^{\,K}\cdot
M^{-1}{\vec r}^{\,L}){\cal L}_{LJ}\Big)\,,
\end{equation}
where the tensor potential is defined as
\begin{equation}
\chi_{{\rm T}}(v,\bar v,x)\equiv -{\cal L}+x^I{\cal L}_I\ ,
\end{equation}
and the matrix $M$ appearing in the inner product is defined as
\begin{equation}
M^{rs}=\frac{1}{4}\Big[{\cal L}_{IJ}({\vec r}^{\,I}\cdot {\vec r}^{\,J})
\delta^{rs}-r^{Ir}{\cal L}_{IJ}r^{Js}\Big],
\end{equation}
with the indices $r,s$ running over three values, and ${\vec r}$
as in \eqref{su2xe}.

Notice that the tensor potential $\chi_T$ is related to the hyperk\"ahler
potential $\chi$ as defined in \eqref{legendre-tr}. In fact, after eliminating
the scalars $x^I$ in terms of $w_I+{\bar w}_I$ by the Legendre transform,
they become the same function, up to a sign,
\begin{equation}\label{chiTchi}
\chi_{\rm T}\Big(v,\bar v, x(w+\bar w,v,\bar v)\Big)=-\chi (v,\bar v, w+\bar w)\,.
\end{equation}
Using the results of the main text \eqref{chiK}, we have
\begin{equation}\label{tp-cmap}
\chi_{\rm T}(v,\bar v,x)=-\frac{v^\flat{\bar v}^\flat}{(r^\flat)^3}
K(\eta_+,\eta_-)\,.
\end{equation}
To make further progress, we list some of the second derivatives of
${\cal L}$.
To facilitate the notation we introduce the scale and $SU(2)$ invariant
variables
\begin{equation}
A^\Lambda=\frac{1}{2(r^\flat)^2}\Big(x^\flat x^\Lambda
+ 2(v^\flat{\bar v}^\Lambda +
{\bar v}^\flat v^\Lambda)\Big)\,.
\end{equation}
Notice that, in relation to the main text, we have that
$2A^\Lambda=\zeta^\Lambda$, as stated before.
With this, and using homogeneity of the
prepotential $F$, we compute the second derivatives
\begin{equation}
{\cal L}_{\Lambda\Sigma}=-\frac{\I}{2r^\flat}\Big(F_{\Lambda\Sigma}-
{\bar F}_{\Lambda\Sigma}\Big),\qquad
{\cal L}_{\flat\Lambda}=-2A^{\Sigma}{\cal L}_{\Lambda\Sigma}\,,
\end{equation}
and
\begin{equation}\label{L00}
{\cal L}_{\flat\flat}=-\frac{\chi_{\rm T}}{(r^\flat)^2}+4A^\Lambda A^\Sigma
{\cal L}_{\Lambda\Sigma}\,.
\end{equation}
These entries are needed to compute the matrix elements of
the matrix $M$. In fact, we need to determine the inverse of $M$ in
\eqref{HIJ}. For that, we first need to find the determinant of $M$, whose
general form was given in \cite{deWit:2006gn}
\begin{equation}
{\rm det}[M]=\frac{1}{3\cdot 4^3}\Big[({\cal L}_{IJ}{\vec
r}^I\cdot{\vec r}^J)^3-{\rm Tr}(PQ)\Big],
\end{equation}
where the matrices $P$ and $Q$ are defined as
\begin{equation}\label{defPQ}
P_{IJ}\equiv {\cal L}_{IK}({\vec r}^K\cdot {\vec r}^L){\cal L}_{LJ}\,,\qquad
Q^{IJ}\equiv ({\vec r}^I\cdot {\vec r}^K){\cal L}_{KL}
({\vec r}^L\cdot {\vec r}^J)\,.
\end{equation}
Defining the quantity
\begin{equation}\label{defY}
Y^\Lambda \equiv \frac{v^\flat}{(r^\flat)^{3/2}}\eta^\Lambda_+\,,
\end{equation}
we can compute
\begin{eqnarray}\label{def-P}
P_{\Lambda\Sigma}&=&\frac{1}{2r^\flat}\Big[(NY)_\Lambda (N{\bar Y})_\Sigma
+(NY)_\Sigma (N{\bar Y})_\Lambda\Big],\nonumber\\
P_{\flat\Lambda}&=&-\frac{1}{r^\flat}\Big[(NY)_\Lambda (N{\bar Y})_\Sigma
+(NY)_\Sigma (N{\bar Y})_\Lambda\Big]A^\Sigma\,,\nonumber\\
P_{\flat\flat}&=&\frac{{\chi}_{\rm T}^2}{(r^\flat)^2}+\frac{4}{r^\flat}(NY)_\Lambda
(N{\bar Y})_\Sigma A^\Lambda A^\Sigma\,,
\end{eqnarray}
where we used the notation $(NY)_\Lambda=N_{\Lambda\Sigma}Y^\Sigma$.
Similarly we find for the matrix elements of $Q$
\begin{eqnarray}
Q^{\Lambda\Sigma}&=&-4(r^\flat)^2\chi_{\rm T}A^\Lambda
A^\Sigma\nonumber\\&& -2r^\flat[(YN{\bar Y})(Y^\Lambda{\bar
Y}^\Sigma+Y^\Sigma{\bar Y}^\Lambda)+Y^\Lambda Y^\Sigma ({\bar
Y}N{\bar Y})+{\bar Y}^\Lambda
{\bar Y}^\Sigma (YNY)]\,,\nonumber\\
Q^{\flat\Lambda}&=&-2\chi_{{\rm T}}(r^\flat)^2A^\Lambda\,,\nonumber\\
Q^{\flat\flat}&=&-{\chi}_{\rm T}(r^\flat)^2\,.
\end{eqnarray}
For the first line we used the identity
\begin{equation}
({\vec r}^\Lambda\cdot {\vec r}^\Sigma)=2r^\flat(Y^\Lambda{\bar
Y}^\Sigma+Y^\Sigma {\bar Y}^\Lambda)+4(r^\flat)^2A^\Lambda A^\Sigma\,,
\end{equation}
which can be proven from the relation
\begin{equation}\label{eta-eta}
\eta^\Lambda_+\eta^\Sigma_-+\eta^\Lambda_-\eta^\Sigma_+=
\frac{1}{2v^\flat{\bar v}^\flat}\big({\vec r}^\flat\times {\vec r}^\Lambda\big)
\cdot\big({\vec r}^\flat\times {\vec r}^\Sigma\big).
\end{equation}
Straightforward computation yields
\begin{equation}
{\rm Tr}(PQ)=-K^3-6|YNY|^2K\,,
\end{equation}
from which one can find the formula for the determinant \eqref{detM},
\begin{equation}\label{detM}
{\rm det}[M]=\frac{1}{32}(YNY)({\bar Y}N{\bar Y})(YN{\bar Y})\,.
\end{equation}

The inverse matrix $M^{-1}$ was also given in \cite{deWit:2006gn}. In
our notation, and using \eqref{defPQ}, it can be rewritten as
\begin{eqnarray}
(M^{-1})_{rs}&=&\frac{32}{{\rm det}[M]}\Big[\big(\chi^2-P_{IJ}({\vec r}^I\cdot
{\vec r}^J)\big)\delta_{rs}+2({\vec r}^I)_r P_{IJ}({\vec r}^J)_s\Big]
\nonumber\\
&=&-\frac{2}{|YNY|^2K(Y,\bar Y)}\Big[\big(K^2+|YNY|^2
\big)\delta_{rs}-({\vec r}^I)_r P_{IJ}({\vec r}^J)_s\Big]\,.
\end{eqnarray}
Finally we compute
\begin{equation}
{\cal H}_{IJ}=\chi_{\rm T}
{\cal L}_{IJ}+\frac{1}{2|YNY|^2}\Big(
(K^2+|YNY|^2)P_{IJ}-(PQ)_I{}^K{\cal L}_{KJ}\Big),
\end{equation}
and define the dilaton as
\begin{equation}
{\e}^\phi=\frac{K(Y,{\bar Y})}{4r^\flat}=-\frac{1}{8(r^\flat)^3}
{\cal L}_{\Lambda\Sigma}
({\vec r}^\flat\times {\vec r}^\Lambda)\cdot({\vec r}^\flat\times {\vec r}^\Sigma)
\,,
\end{equation}
with the same normalization as in \cite{Rocek:2005ij}. Notice that
this coincides with the dilaton given in \eqref{phichi} when we identify
$\phi=2U$. We then find for
the components
\begin{eqnarray}
{\cal H}_{\flat\flat}&=&-8{\e}^\phi\Big({\e}^\phi
-({\cal N}+{\bar {\cal N}})_{\Lambda \Sigma}A^\Lambda A^\Sigma\Big),
\nonumber\\
{\cal H}_{\flat\Lambda}&=&-4{\e}^{\phi}A^\Sigma
({\cal N}+{\bar {\cal N}})_{\Sigma\Lambda}\,,\nonumber\\
{\cal H}_{\Lambda\Sigma}&=&2{\e}^\phi
({\cal N}+{\bar {\cal N}})_{\Lambda \Sigma}\,.
\end{eqnarray}
This matches precisely with the matrix ${\cal T}_{IJ}$ in \eqref{defT}
up to an overall normalization factor which is exactly the same as in
\cite{Rocek:2005ij}. One can repeat this analysis for the other terms
in the Lagrangian, but these are fixed by supersymmetry, and moreover
in \cite{Rocek:2005ij} they were shown to be correctly reproduced in the
$SU(2)$ gauge $v^\flat=0$.

This concludes the proof that the superspace Lagrangian \eqref{SS-lagr}
describes the $c$-map Lagrangian \eqref{cmap-lagr}, and validates the
``covariant $c$-map'' formulae in Section \ref{sec-covariant-cmap}.

\bibliography{combined}
\bibliographystyle{utphys}

\end{document}